\documentclass[onecolumn]{aa} 
\usepackage{graphicx}
\usepackage{txfonts}
\begin{document}
\title{The penetration of FUV radiation into molecular clouds}


\author{Javier R. Goicoechea\inst{1}
\and Jacques Le Bourlot\inst{2}
       }

\offprints{J.R. Goicoechea or J. Le Bourlot}

\institute{LERMA--LRA, UMR 8112, CNRS, Observatoire de Paris and \'Ecole Normale
Sup\'erieure, 24 Rue Lhomond, 75231 Paris cedex 05, France.\\
\email{javier@lra.ens.fr}
\and
LUTH, UMR 8102 CNRS, Universite Paris 7 and Observatoire de Paris,  Place J. Janssen
92195 Meudon, France.\\
\email{Jacques.Lebourlot@obspm.fr}
             }

   \date{Received 2006 July 26; accepted 2007 January 30}

 
  \abstract
    {FUV radiation  strongly affects the physical and chemical state of 
     molecular clouds, from protoplanetary disks to entire galaxies.}  
    {The solution of the FUV radiative transfer equation can be complicated if the most
    relevant radiative processes such us dust scattering and gas line absorption 
    are included, and  have realistic (non--uniform)
    properties, i.e. if optical properties are depth dependent.}
    {We have extended the \textit{spherical harmonics} method to solve for the FUV radiation field
    in externally or internally illuminated clouds taking into account gas 
    absorption and coherent, nonconservative and anisotropic scattering by dust grains. 
    The new formulation has been implemented in the \textit{Meudon PDR code} and thus it will be
    publicly available.} 
    {Our formalism allows us to consistently include:
    $(i)$ varying dust populations and $(ii)$ gas lines in the FUV radiative transfer.
    The FUV penetration depth rises for increasing dust albedo and anisotropy of the scattered radiation
    (e.g. when grains grow towards cloud interiors). }
    {Illustrative models of illuminated clouds where only the dust populations are varied
     confirm earlier predictions for the FUV penetration in  diffuse clouds (A$_V$$<$1). 
     For denser and more embedded sources (A$_V$$>$1) we show that
     the FUV radiation field inside the
     cloud can differ by orders of magnitude  depending on the grain properties and growth.
     Our models reveal significant differences regarding the resulting physical and
     chemical structures for steep vs. flat extinction curves towards molecular clouds. 
     In particular, we show that the photochemical and thermal gradients can be very 
     different depending on grain growth. Therefore, the assumption of uniform 
     dust properties and averaged extinction curves can be a crude approximation to determine
     the resulting  scattering properties, prevailing chemistry and atomic/molecular abundances
     in  ISM clouds or protoplanetary disks.}

  
   \authorrunning{Goicoechea \& Le Bourlot} 
   \titlerunning{The penetration of FUV radiation} 

   \maketitle

\section{Introduction}

Far--UV (FUV) radiation ($h\nu$ $<$13.6~eV)  strongly affects the physical and chemical
state of dusty molecular clouds in many evolutionary stages:
from star forming regions (\cite{Leq81,Stu88,Bal98}) and protoplanetary disks (\cite{Joh98,Aik02}), 
to  circumstellar  envelopes around evolved stars (\cite{Hug82,Hab96})
 and supernova remnants (\cite{Shu79,Che94}).
Thus, the accurate knowledge of the intensity of the FUV radiation field as a function 
of cloud depth is of crucial importance in a plethora of astrophysical environments.
Penetration of FUV radiation strongly depends on
dust grains properties  through the scattering of  photons,
but it also depends on the gas properties (chemical composition, density, etc.)
through the absorption 
of hundreds of discret electronic lines from the most abundant species
(H,~H$_2$,~and~CO). This proccess is, in addition, an efficient excitation mechanism for molecular
species  (\cite{Bla87, Ste89}). Gas absorption lines reach
extremely large opacities and, due to saturation, they can be very broad and  fully absorb the 
FUV continuum. 

The so called \textit{spherical harmonics} method, in which the specific
intensity of the FUV radiation field is expanded into series of Legendre polynomials, is
an efficient way to solve the plane--parallel radiative transfer equation
if gas opacity is neglected and if dust grains have uniform optical properties,
e.g. the same extinction cross--section, albedo and scattering phase 
function (\cite{Fla80, Rob83}). 
Nevertheless, astronomical observations over the full spectral domain
show a more complex scenario, where dust grain populations evolve 
depending on the environmental conditions from polycyclic
aromatic hydrocarbons (PAHs) and very small grains (VSGs) to 
bigger grains (BGs) likely formed by accretion or coagulation 
(\cite{Bou88,Des90,Job92,Dra03,Dar05}).
Also, the average extinction law (\textit{e.g.,} \cite{Car89}) 
is based on observations toward low--extinction line of sights ($A_V\lesssim5$), and it
has been questioned by recent observations toward more embedded regions
($A_V\geq15$). A better knowledge of the extinction properties at large A$_V$ is critical. 
In particular, there is evidence that the reddening curve
tends to flatten at high extinction depths (\cite{Moo05}),  
consistent with grain growth  and dust processing along the line of sight.
Therefore, the attenuation of  FUV radiation will dramatically   
depend on the (generally poorly understood)  grain composition
and optical properties that, of course, are likely to change from source to source
according to the  interstellar (ISM) and circumstellar (CSM) dust  life--cycle. 
In addition to this dust--shielding, self--shielding 
 through gas line absorption  can result in an
efficient protection of H$_2$ and CO, and the starting point
of a rich chemistry even in irradiated media
such as protoplanetary disks, translucent clouds, starbursts galaxies or, more generally, 
\textit{photodissociation regions} (PDRs; see \cite{Hol97} for a review). 

 The \textit{spherical harmonics} method also has been implemented 
to study the radiative transfer and dust extinction in galaxies as a whole by associating the
source function with the emissivity of a given distribution of stars through the galaxy
(\cite{Bar95}; \cite{Bae01}). Uniform  grain properties and the absence of gas line 
absorption are assumed. 
For unidimensional problems, the \textit{spherical harmonics} method
 is found to be by far the most efficient way to solve for the radiative transfer equation
compared to Monte Carlo or ray tracing techniques (\cite{Bae01}).

The detailed information provided by high angular resolution observations 
(e.g. \cite{Ger05,Goi06}), revealing fine differences even between
similar sources, should be followed by a sophistication in the
radiative transfer modeling.
Inclusion of gas (discrete line absorption) and varying grain populations
(e.g. different extinction curves)
as a function of  cloud depth   requires a 
modification of the original method (\cite{Fla80}; \cite{Rob83}).
In this work we present an extension  of the 
\textit{spherical harmonics}
method for a radiative transfer equation with depth dependent coefficients
in plane--parallel geometry.
We used this method to solve for the  radiation
field in  illuminated clouds at wavelengths longer than Lyman cut--off at $\sim$912~\AA$\,$
taking into account  gas absorption and scattering by dust grains.
The method can also include the source function for
embedded emission of photons, and therefore it can explicitly take
into account any source of internal radiation. 

In Secs.~2 and 3 we present the formulation of the method while in 
Secs.~4 and 5 we show several astrophysical applications  to understand
the role of FUV penetration for the photochemistry of molecular clouds. 
In particular, we present a few examples including  H Lyman lines,  
H$_2$ electronic transitions within the Lyman and Werner bands and CO electronic
transitions together with varying dust properties.
The penetration of FUV radiation for the typical conditions prevailing in a diffuse cloud 
(such us $\zeta$~Ophiuchi) and in  higher extinction objects  (such as the Orion~Bar or a strongly
illuminated protoplanetary disk) are discussed.

\section{The equation of radiative transfer with variable coefficients}

The specific intensity of radiation, $I_{\lambda}(s,\mu)$, 
in plane-parallel geometry is a solution of the radiative transfer equation:
\begin{equation}
\mu\frac{\partial I_{\lambda}(s,\mu)}{\partial s}=-[\alpha_{\lambda}(s)+
\sigma_{\lambda}(s)]\, 
I_{\lambda}(s,\mu)+\frac{\sigma_{\lambda}(s)}{2}\,\,\int_{-1}^{+1}R_{\lambda}(s,\mu,\mu')\, 
I_{\lambda}(s,\mu')\, d\mu'+ j_{\lambda}(s)
\label{eq-etr1}
\end{equation}
where the spatial scale $s$ and the angle $\theta=cos^{-1}\mu$ are the independent
variables and where the dependence of quantities on wavelength $\lambda$ and
on $s$ has been explicitly written. In the most general problem, 
$\alpha_{\lambda}(s)=\alpha_{\lambda}^{g}(s)+\alpha_{\lambda}^{d}(s)$ 
is the line--plus--continuum absorption coefficient, 
$\sigma_{\lambda}(s)$  is the dust scattering coefficient, 
$j_{\lambda}(s)$ 
is the emission coefficient of any source of internal radiation 
and $R_{\lambda}(s,\mu,\mu')$ is the angular redistribution function
(we assume that the radiation field has azimuthal symmetry
about normal rays). In this work, the opacity is due to 
coherent (no energy redistribution in the scattered photons), 
nonconservative (a fraction of photons are absorbed), anisotropic scattering by dust grains 
as well as to gas line absorption, that is: 
\begin{equation}
 d\tau=-(\alpha_{\lambda} + \sigma_{\lambda})\,ds
 \label{eq-total-tau} 
\end{equation}
(note that $\tau$ increases toward the decreasing direction of $s$) and the radiative transfer 
Eq.~(\ref{eq-etr1}) gets the more familiar form
\begin{equation}
\mu\frac{\partial I_{\lambda}(\tau,\mu)}{\partial \tau} 
= I_{\lambda}(\tau,\mu) -  \frac{\omega_{\lambda}(\tau)}{2}\,
 \int_{-1}^{+1} R_{\lambda}(\tau,\mu,\mu')\, I_{\lambda}(\tau,\mu')\, d\mu'-S^{*}_{\lambda}(\tau,\mu)
=
I_{\lambda}(\tau,\mu) - S_{\lambda}(\tau,\mu) 
\label{eq-etr2}
\end{equation}
where $\omega_{\lambda}=\frac{\sigma_{\lambda}}{\alpha_{\lambda}+\sigma_{\lambda}}$ 
is a new effective albedo (the dust scattering cross--section over
the total dust+gas extinction cross--section) which tends to the pure dust albedo
for wavelengths free of lines, but tends to 0 (true gas absorption) at the line cores.
Intermediate values are found in the line wings.
$S^{*}_{\lambda}=\frac{j_{\lambda}}{\alpha_{\lambda}+\sigma_{\lambda}}$ is
the source function for the true emission by "embedded photon sources". 
In the following we assume that $S_{\lambda}^{*}$=0. Thus we ignore dust thermal emission
(negligible in the FUV for ISM clouds) or any other source of internal illumination.
Hence, our source function only corresponds to the external illumination
photons scattered by dust grains.
However, inclusion of $S_{\lambda}^{*}$ in our method is trivial. The interested reader
is refererred to Appendix~\ref{ap_embedded}.

\begin{figure*}[t]
    \centering \includegraphics[width=0.7\hsize{}]{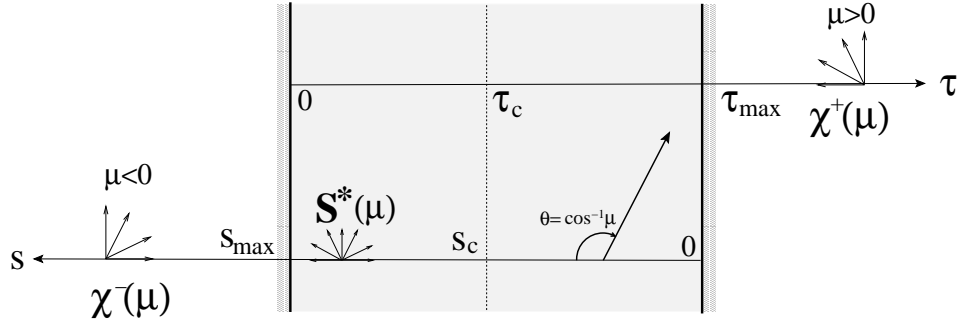}
    \caption{Adopted geometry and sign conventions for a cloud 
    with embedded sources of photons $S^*(\mu)$ and illuminated at both surfaces
    by $\chi^{\pm}(\mu)$.}
    \label{fig:geo}
\end{figure*}

The cloud extends from $\tau=0$ to $\tau=\tau_{max}$
with a possibility that $\tau_{max}=\infty$. Boundary conditions
require $I_\lambda(\tau,\mu)$ to match the incident intensity at $\tau=0$
and $\tau=\tau_{max}$. Note the implicit sign convention on $\mu$:
$\theta=\pi$ points towards positive values of $\tau$, that is
$\mu=-1$ for a ray perpendicular to the cloud and penetrating into
it from $\tau=0$ (see Fig.~\ref{fig:geo}). Thus, boundary conditions specify  
$\chi^-(\mu)=I_{\lambda}(0,\mu)$ ($\mu<0$) and 
$\chi^+(\mu)=I_{\lambda}(\tau_{max},\mu)$ ($\mu>0$), where $\chi^{\pm}(\mu)$ are the
illuminating radiation fields reaching both cloud surfaces (of course they can be different).

Compared to other works where the \textit{spherical harmonics} method has been applied
to solve for the FUV radiation field  (e.g. \cite{Fla80}; \cite{Rob83};  \cite{Bar95};
\cite{Bae01}; \cite{Pet06}), the optical properties  
in the radiative transfer equation  (e.g., effective albedo and asymmetry parameter)
are  wavelength-- and  cloud depth--dependent for the first time.

\section{The spherical harmonics method for line and continuum transfer}
\label{sec-method}

\subsection{The $P_L$ approximation\label{subsec-Pl}}

In this method, the angular dependence of the radiation field $I(\tau,\mu)$ 
is expanded in a truncated series of Legendre polynomials $P_{l}(\mu)$ which form
a complete orthogonal set within the range (-1,1) in which $\mu$ varies:
\begin{equation}
I(\tau,\mu)=\sum_{l=0}^{L}(2l+1)\, f_{l}(\tau)\, P_{l}(\mu)
\label{eq-expI}
\end{equation}
where the dependence on $\lambda$ is no longer shown. 
In the following
sections we show that the mean intensity of the radiation field at
each depth point $J(\tau)$ has the simple form 
$J(\tau)=f_0(\tau)$, 
i.e. the first coefficient of the expansion in Eq.~(\ref{eq-expI}), which is
often the only quantity needed for the integration of radiation field--dependent
physical parameters (e.g. photoionization and photodissociation rates). 
This is one of the reasons why the method is so attractive.
However, a large number of expansion terms  
has to be used in order to correctly sample the angular dependence of the
radiation field, we typically use $L+1=2M=20$ (note that dust scattering can 
be highly anisotropic at the considered wavelengths). 

If the grain scattering phase function $p(\tau,cos\,\Theta)$ only depends
on the angle $\Theta$ between the incident and scattered radiation,
$R(\tau,\mu,\mu')$ can also be expanded 
(see e.g., \cite{Cha60}; \cite{Rob83}) as:
\begin{equation}
R(\tau,\mu,\mu')=\sum_{l=0}^{L}(2l+1)\,\sigma_{l}(\tau)\, P_{l}(\mu)\, P_{l}(\mu')
\label{eq-expR}
\end{equation}
in terms of the $\sigma_{l}(\tau)$ coefficients
of the Legendre expansion of  $p(\tau,cos\,\Theta)$:
\begin{equation}
p(\tau,cos\,\Theta)=\sum_{l=0}^{L}(2l+1)\,\sigma_{l}(\tau)\, P_{l}(cos\,\Theta)
\label{eq-expR2}
\end{equation} 
The standard model of scattering by interstellar grains (\cite{Hen41})
assumes the simple scattering phase function:
\begin{equation}
p(cos\,\Theta)= \frac {1-g^2} {(1+g^2-2g\,cos\,\Theta)^{3/2}}
\label{eq-HGfun}
\end{equation} 
which can be also expanded  in Legendre polynomials in terms of the 
"$g$--asymmetry parameter"  ($=<cos\,\Theta>$) i.e., the mean angle of the
scattered radiation ($g=1/2 \int_{-1}^{+1} \mu\,p(\mu)\,d\mu$, with $\mu=cos\,\Theta$).
 Here we adopt a $\tau$-dependent \textit{Henyey--Greenstein phase function} 
(other phase functions can be used if they can also be expanded). 
Therefore
we  write:
\begin{equation}
p(\tau,cos\,\Theta)= \sum_{l=0}^{L}(2l+1)\,g_{l}(\tau)\, P_{l}(cos\,\Theta)
\label{eq-HGfun2}
\end{equation} 
where $g_l(\tau)=\sigma_l(\tau)$ and $g_0(\tau)=1$.
Thus, the angular redistribution function $R(\tau,\mu,\mu')$ has two obvious
limiting cases, $g(\tau)=0$ (isotropic scattering) and
$g(\tau)=\pm1$  
with $R(\tau,\mu,\mu')=\delta(\mu\mp\mu')$ (pure  backward or forward  scattering).
 
Substitution of Eqs.~(\ref{eq-expI}) and (\ref{eq-expR}) into
the transfer equation (\ref{eq-etr2}) and using appropriate recurrence formulae
leads to the finite ($L+1$) set
of coupled, linear, first order differential equations in the 
unknown $f_l(\tau)$ coefficients, with $l=0,...,L$.
\begin{equation}
lf'_{l-1}(\tau)+(l+1)f'_{l+1}(\tau)=(2l+1)\,[1-\omega(\tau)\sigma_{l}(\tau)]
\, f_{l}(\tau)
\label{eq-system}
\end{equation}
where $f'=\partial f /\partial \tau$. We recall that compared to
Roberge (1983) this is \textbf{not} a constant coefficient
equation 
so numerical integration is necessary.
In the "$P_L$ approximation" a sufficiently large odd\footnote{For even values 
of $L$, $\mathbf{A}$ is singular (e.g. Roberge 1983)}  $L$ value has to be chosen
to obtain an accurate solution of the problem.
The system (\ref{eq-system}) can be written as:
\begin{equation}
\mathbf{f'}(\tau)=\mathbf{A^{-1}\mathnormal{(\tau)\,}f}(\tau)
\label{eq-system2}
\end{equation}
with:
\begin{equation}
\mathbf{A}(\tau)=\left(\begin{array}{cccccc}
0 & h_{0}^{-1}\\
h_{1}^{-1} & 0 & 2h_{1}^{-1}\\
 & 2h_{2}^{-1} & 0 & \ddots\\
 &  & 3h_{3}^{-1} &  & (L-1)h_{L-2}^{-1}\\
 &  &  & \ddots & 0 & Lh_{L-1}^{-1}\\
 &  &  &  & Lh_{L}^{-1} & 0\end{array}\right)
\end{equation}
where:
\begin{equation}
h_{l}(\tau)=(2l+1)\,(1-\omega(\tau)\sigma_{l}(\tau)) \hspace{1cm} 
\end{equation} 

In summary, we have to solve for a linear boundary value problem with non 
constant coefficients with the additional difficulty of huge variations
of the total opacity\footnote{We also developed the formalism to solve 
Eq.~(\ref{eq-system2}) through \textit{finite differences} (\cite{Ash95}). 
For only dust continuum transfer, results are almost identical (within $\sim$0.1\%)
to those obtained with the \textit{spherical harmonics} method (which is $\sim$2 times faster).
 However, when  line absorption is included, the \textit{finite difference} numerical solution always oscillate
at the core of saturated lines and no optimal solution is found.} 
within small variations in the wavelength and
cloud position grids, e.g. from $\lambda$ in a saturated line center 
($\tau_{\lambda}\sim$10$^7$)
to  $\lambda$ in an adjacent (line free) continuum region ($\tau_{\lambda}\sim$10).
 In the following, we show an extension of the spectral method of Flannery et al. (1980)
and Roberge (1983) to solve for the FUV radiative transfer. 

\subsection{The eigenvalues solution}

\subsubsection{Numerical solution}\label{sub-sec-numsol}

The  $\mathbf{A}^{-1}(\tau)$ matrix has $L+1=2M$ eigenvalues which are
real, non-zero and non-degenerate and which occur in positive-negative
pairs, see Appendix~A of Roberge (1983). Using  a similar notation as Roberge,
let $k_{m}(\tau),\, m=\pm1,\cdots,\pm M$ be the 
eigenvalues verifying $k_{-m}(\tau)=-k_{m}(\tau)$, and $\mathbf{R(\tau)}$ be the 
matrix of eigenvectors, that is:
\begin{equation}
\sum_{j}\mathbf{A}^{-1}(\tau)_{lj}\, R_{jm}(\tau)=k_{m}(\tau)\, R_{lm}(\tau)
\label{eq-Asis}
\end{equation}
which also verifies the $R_{l,-m}(\tau)=(-1)^{l}\, R_{lm}(\tau)$ relation. 
The depth--dependence of the eigenvalues $k_{m}(\tau)$ and eigenvectors
$R_{lm}(\tau)$ complicates the solution of the problem compared to the
(only dust) problem with uniform optical grain  properties.
The computation of $k_{m}(\tau)$ and $R_{lm}(\tau)$ is given in Appendix~\ref{ap_diag}.

The $\mathbf{R}(\tau)$ matrix of eigenvectors can still be used to define an auxiliary
set of variables $\mathbf{y}(\tau)=\mathbf{R^{-1}}(\tau)\,\mathbf{f}(\tau)$, or:
\begin{equation}
f_{l}(\tau)=\sum_{-M}^{-1}R_{lm}(\tau)\, y_{m}(\tau)+\sum_{1}^{M}R_{lm}(\tau)\, 
y_{m}(\tau)
\end{equation}
so that
\begin{equation}
\mathbf{f}'=\mathbf{A}^{-1}\,\mathbf{R}\, \mathbf{y}
\end{equation}
Therefore, in terms of 
the new $\mathbf{y}(\tau)$ variables, Eq.~(\ref{eq-system2})
can be rewritten as:
\begin{equation}
\mathbf{y'}=\mathbf{R}^{-1}\,\mathbf{A}^{-1}\,\mathbf{R}\,\mathbf{y}
-\mathbf{R}^{-1}\,\mathbf{R}'\, \mathbf{y}
=\mathbf{K}\,\mathbf{y}-\mathbf{R}^{-1}\,\mathbf{R}'\,\mathbf{y}
\label{eq-neweq}
\end{equation}
To write Eq.~(\ref{eq-neweq}) we have used the fact that 
$(\mathbf{R^{-1}A^{-1}R})_{lm}=k_l\,\delta_{lm}$
and thus $\mathbf{K}(\tau)$ is a diagonal matrix with the $k_{m}(\tau)$ eigenvalues
of $\mathbf{A^{-1}(\tau)}$ on its diagonal.
The fact that $\mathbf{R}'(\tau)\neq0$ adds the last matrix term 
in Eq.~(\ref{eq-neweq}) due to the depth--dependence of the coefficients.
This term is neglected in Le Petit et al. (2006). However, $\mathbf{R}'$
is not null neither when the grain optical
properties depend on the cloud depth (even if gas is neglected) nor 
when gas line absorption is included (even if grain properties are uniform).
Unfortunately, the system of 
Eqs.~(\ref{eq-neweq}) is uncoupled 
only if the $\mathbf{R}^{-1}\,\mathbf{R}'\, y$ term is null (as in Roberge 1983), 
otherwise more manipulations are required to solve the problem consistently. 
If we define 
$\mathbf{Q}=\mathbf{R}^{-1}\mathbf{R}'\,\mathbf{y} = - \mathbf{L}\,\mathbf{y}\,$, 
then Eq.~(\ref{eq-neweq}) can be simply written as:
\begin{equation}
y'_{m}=k_{m}(\tau)\, y_{m} + [\mathbf{L\,y}]_m(\tau)
\label{eq-new-eq}
\end{equation} 
for $m=\pm1,...,\pm M$. In order to solve this particular problem we  turn the system
of differential equations~(\ref{eq-new-eq}) into an integral problem.
To do that we first introduce the following integral equation: 
\begin{equation}
y_{m}(\tau)=e^{a_m(\tau)} \left[ C_m + \int_{\tau_{m}}^{\tau}\, e^{-a_m(t)}\,
 [\mathbf{L\,y}]_m(t)\, dt \right]
\label{eq-integral-form}
\end{equation}
where  $a_m(\tau)$ is an arbitrary function so that $a_m(\tau_m)$=0. The system of
Eqs.~(\ref{eq-integral-form}) represents a general set of integral equations
that verify $y_m(\tau_m)=C_m$ (to be found from the boundary conditions). 
If a given function $y_m$ is a solution of the above
equation, by taking its derivative with respect to $\tau$ one gets:
\begin{equation}
y'_{m}(\tau)=a'_{m}(\tau)y_{m}(\tau) + [\mathbf{L\,y}]_m(\tau)  
\end{equation}
which means that $y_m$ as defined in Eqs.~(\ref{eq-integral-form}) is also
a solution of the original system of differential Eqs.~(\ref{eq-new-eq}) if and
only if $a'_m(\tau)=k_m(\tau)$. Therefore, $a_m(\tau)=\int_{\tau_{m}}^{\tau}\,k_m(t)\,dt$.
This demonstration shows that the $k_m$ eigenvalues of $\mathbf{A^{-1}}$ (and no others)
are the right exponential factors that  do attenuate the radiation field, which
is consistent with the original problem described by Eqs.~(\ref{eq-system2}). 
 In the present work we solve Eqs.~(\ref{eq-integral-form}) with an iterative scheme$^2$ and
thus compute:
\begin{equation}
y_{m}^{(n+1)}(\tau)=e^{\int_{\tau_{m}}^{\tau}\,k_{m}(t)\,dt} 
\left[ C_{m}^{(n+1)} + \int_{\tau_{m}}^{\tau}\, e^{-\int_{\tau_{m}}^{t}\,k_{m}(t')\,dt'} \,
[\mathbf{L\,y}^{(n)}]_m(t)\,  dt \right]
\label{eq-iterate}
\end{equation}
by using an appropriate (physical) initial guess for $y_{m}^{(n)}$, where $n$ is the
iteration step.
This iterative procedure shows that the solution if forced, at any step, by the
exponential factor $e^{\int_{\tau_{m}}^{\tau}\,k_{m}(t)\,dt}$ to follow the behavior
dictated by the "true" eigenvalues of the problem (i.e. those of the
original coupling matrix $\mathbf{A^{-1}}$) that are known before
the iteration procedure is started.
In Appendix~\ref{ap_error} we give details on the error bound associated with the iterative
scheme and we show that the numerical solution derived for the FUV radiation
field correctly satisfies the original system of Eqs.~(\ref{eq-system2}).


At each iteration step we have to compute the 
integration constants $C_{m}$ by a convenient selection of $\tau_m$.
To ensure that only exponentials with  negative arguments appear,
it is necessary to set $\tau_{m}=0$ for $m<0$ and $\tau_{m}=\tau_{max}$ for $m>0$.
In order to have  easier to read equations, we now  introduce
some convenient notations: 
\begin{equation}
E_{m}^{-}(\tau)=\exp\left(\int_{0}^{\tau}k_{m}(t)\, dt\right)\,\,\,(m<0)
\,\hspace{1cm}{\rm or}\hspace{1cm}\,E_{m}^{-}(\tau)=\exp\left(-\int_{0}^{\tau}k_{m}(t)\, 
dt\right)\,\,\,(m>0)
\end{equation}
Note that $E_{-m}^{-}(\tau)=E_{m}^{-}(\tau)$, and $E_{m}^{-}(0)=1$. We also define:
\begin{equation}
E_{m}^{+}(\tau)=\exp\left(\int_{\tau}^{\tau_{max}}k_{m}(t)\, dt\right)\,\,\,(m<0)
\,\hspace{1cm}{\rm or}\hspace{1cm}\, E_{m}^{+}(\tau)=\exp\left(-\int_{\tau}^{\tau_{max}}
k_{m}(t)\, dt\right)\,\,\,(m>0)
\end{equation}
with $E_{m}^{+}(\tau_{max})=1$ and $E_{m}^{+}(\tau)\times 
E_{m}^{-}(\tau)=E_{m}^{+}(0)=E_{m}^{-}(\tau_{max})$.
Using the above notations, we have:
\begin{equation}
y_{m}(\tau)=E_{m}^{-}(\tau)\, C_{m}-\int_{0}^{\tau}\frac{E_{m}^{-}(\tau)}
{E_{m}^{-}(t)}\,q_{m}(t)\, dt\qquad (m<0)
\end{equation}
\begin{equation}
y_{m}(\tau)=E_{m}^{+}(\tau)\, C_{m}+\int_{\tau}^{\tau_{max}}
\frac{E_{m}^{+}(\tau)}{E_{m}^{+}(t)}\,q_{m}(t)\,
 dt\qquad (m>0)
\end{equation}
Note the change of sign in the second equation due to the inversion
of $\int_{\tau_{max}}^{\tau}$. To further simplify these expressions, we define:
\begin{equation}
D_{m}^{-}(\tau)=\int_{0}^{\tau}\frac{E_{m}^{-}(\tau)}{E_{m}^{-}(t)}\,q_{m}(t)\, dt\qquad (m<0)
\label{eq-D-}
\end{equation}
\begin{equation}
D_{m}^{+}(\tau)=\int_{\tau}^{\tau_{max}}\frac{E_{m}^{+}(\tau)}{E_{m}^{+}(t)}\,
q_{m}(t)\, dt\qquad (m>0)
\label{eq-D+}
\end{equation}
which satisfy $D_{m}^{-}(0)=0$ and $D_{m}^{+}(\tau_{max})=0$. Therefore,
the $\mathbf{y}(\tau)$ variables are finally written  compactly as:
\begin{equation}
y_{m}(\tau)=E_{m}^{-}(\tau)\, C_{m}-D_{m}^{-}(\tau)\qquad (m<0)
\end{equation}
\begin{equation}
y_{m}(\tau)=E_{m}^{+}(\tau)\, C_{m}+D_{m}^{+}(\tau)\qquad (m>0)
\end{equation}
and the original $f_l(\tau)$ terms in the Legendre expansion of
the radiation field  $I(\tau,\mu)$ are then given by:
\begin{equation}
f_{l}(\tau)=\sum_{m=-M}^{-1}R_{lm}(\tau)\,\left(C_{m}\, 
E_{m}^{-}(\tau)-D_{m}^{-}(\tau)\right)
+\sum_{m=1}^{M}R_{lm}(\tau)\,\left(C_{m}\, E_{m}^{+}(\tau)+D_{m}^{+}(\tau)\right).
\label{eq-final_f}
\end{equation}

\subsubsection{Boundary conditions: Clouds with two sides illumination}

We  consider a unidimensional plane--parallel cloud of finite size
with an external radiation field at both cloud surfaces ($\tau=0$ and
$\tau=\tau_{max}$) defined by  $\chi^-(\mu)$ and $\chi^+(\mu)$ 
respectively (see Fig.~\ref{fig:geo}). From Eq.~(\ref{eq-final_f}) we have:
\begin{equation}
f_{l}(0)=\sum_{m=-M}^{-1}R_{lm}(0)\, 
C_{m}+\sum_{m=1}^{M}R_{lm}(0)\,\left(C_{m}\, E_{m}^{+}(0)+D_{m}^{+}(0)\right)
\end{equation}
\begin{equation}
f_{l}(\tau_{max})=\sum_{m=-M}^{-1}R_{lm}(\tau_{max})\,\left(C_{m}\,
E_{m}^{-}(\tau_{max})-D_{m}^{-}(\tau_{max})\right)+
\sum_{m=1}^{M}R_{lm}(\tau_{max})\, C_{m}.
\end{equation}
At the $\tau=0$ side, the solution must match,
 at each $\lambda$, the incoming radiation 
field with $\mu<0$, i.e. $I(0,\mu)=\chi^-(\mu)$, with
\begin{equation}
I(0,\mu)=\sum_{l=0}^{L}(2l+1)\, f_{l}(0)\, P_{l}(\mu) \hspace{2cm} {\rm or:}
\label{eq:_BC1}
\end{equation}
\begin{equation}
I(0,\mu)=\sum_{m=-M}^{-1}C_{m}\,\sum_{l=0}^{L}(2l+1)\, R_{lm}(0)\, P_{l}(\mu)
+\sum_{m=1}^{M}\left(C_{m}\, E_{m}^{+}(0)+D_{m}^{+}(0)\right)\,
\sum_{l=0}^{L}(2l+1)\, R_{lm}(0)\, P_{l}(\mu).
\end{equation}
At the $\tau=\tau_{max}$ side, the solution must match, at each $\lambda$, the 
incoming radiation field with $\mu>0$, i.e. $I(\tau_{max},\mu)=\chi^+(\mu)$, with:
\begin{equation}
I(\tau_{max},\mu)=\sum_{l=0}^{L}(2l+1)\, f_{l}(\tau_{max})\, 
P_{l}(\mu) \hspace{2cm} {\rm or:}
\label{eq:_BCN}  
\end{equation}
\begin{equation}
I(\tau_{max},\mu)=\sum_{m=-M}^{-1}\left(C_{m}\, E_{m}^{-}(\tau_{max})-D_{m}^{-}
(\tau_{max})\right)\,\sum_{l=0}^{\infty}(2l+1)\, R_{lm}(\tau_{max})\, P_{l}(\mu)
+\sum_{m=1}^{M}C_{m}\,\sum_{l=0}^{L}(2l+1)\, R_{lm}(\tau_{max})\, P_{l}(\mu).
\end{equation}
Nevertheless, since the order $L$ of the expansions is finite,
the boundary conditions  $I(0,\mu)=\chi^-(\mu)$ and $I(\tau_{max},\mu)=\chi^+(\mu)$
can not be satisfied at all $\mu$ angles. In this work we use 
\textit{Marck's\footnote{See e.g., Sen \& Wilson (1990) for a different choice of boundary conditions.} conditions}
that require $I(0,\mu<0)$ and $I(\tau_{max},\mu>0)$ to match the incident
radiation fields at  $L+1=2M$ strategic angles $\mu_i$ given by $P_{L+1}(\mu_i)=0$, 
that is, the roots of the Legendre polynomial of degree $L+1$.
Note that in these $\mu_{\pm i}$ ($i=\pm 1,...,\pm M$) angles,
the solution of the  radiation field $I(\tau,\mu_i)$ is "exact".

To further simplify the boundary conditions relations, we now define:
\begin{equation}
T_{im}(0,\mu_{i})=\sum_{l=0}^{L}(2l+1)\, R_{lm}(0)\, P_{l}(\mu_{i})\quad (\mu_{i}<0)
\end{equation}
\begin{equation}
T_{im}(\tau_{max},\mu_{i})=\sum_{l=0}^{L}(2l+1)\, R_{lm}(\tau_{max})\, P_{l}(\mu_{i})\quad (\mu_{i}>0)
\end{equation}
which gives:
\begin{equation}
I(0,\mu_{i})=\sum_{m=-M}^{-1}C_{m}\, T_{im}(0,\mu_{i})+\sum_{m=1}^{M}\left(C_{m}\, 
E_{m}^{+}(0)+D_{m}^{+}(0)\right)\, T_{im}(0,\mu_{i})
\end{equation}
\begin{equation}
I(\tau_{max},\mu_{i})=\sum_{m=-M}^{-1}\left(C_{m}\, E_{m}^{-}(\tau_{max})-D_{m}^{-}
(\tau_{max})\right)\, T_{im}(\tau_{max},\mu_{i})+\sum_{m=1}^{M}C_{m}\, T_{im}(\tau_{max},\mu_{i})
\end{equation}
Therefore, the desired $C_m$ constants at each iteration step
are solutions of the $2M\times2M$ linear system ($m=0$ excluded):
\begin{equation}
\sum_{\begin{array}{c}
m=-M\end{array}}^{M}B_{im}\, C_{m}=H_{i}
\label{eq-Bound_C}
\end{equation}
with the $B_{im}$ coefficients as define in Table~\ref{Tab_B_im}, and where
\begin{equation}
H_{i}=\left\{ \begin{array}{lc}
I^{-}(0,\mu_{i})-\sum_{m=1}^{M}D_{m}^{+}(0)\, T_{im}(0,\mu_{i}) & \quad (\mu_{i}<0)\\
I^{+}(\tau_{max},\mu_{i})+\sum_{m=-M}^{-1}D_{m}^{-}(\tau_{max})\, T_{im}(\tau_{max},\mu_{i}) 
& \quad (\mu_{i}>0)\end{array}\right.
\end{equation}
\begin{table}[!htbp]
\begin{centering}
\begin{tabular}{|c||c|c|}
\hline 
$B_{im}=$&
$m<0$&
$m>0$\tabularnewline
\hline
\hline 
$\mu_{i}<0$&
$T_{im}(0,\mu_{i})$&
$T_{im}(0,\mu_{i})\, E_{m}^{+}(0)$\tabularnewline
\hline 
$\mu_{i}>0$&
$T_{im}(\tau_{max},\mu_{i})\, E_{m}^{-}(\tau_{max})$&
$T_{im}(\tau_{max},\mu_{i})$\tabularnewline
\hline
\end{tabular}\par
\end{centering}
\caption{$B_{im}$ coefficients for the two sides illumination
boundary conditions in Eq.~(\ref{eq-Bound_C})
\label{Tab_B_im}}
\end{table}


For semi--infinite clouds ($\tau_{max}$=$\infty$) with only one side illumination
at $\tau$=0 ($\mu_i<0$), 
boundary conditions have to be modified to take into account
the \textit{no radiation} condition at  $\tau$=$\infty$ ($\mu_i>0$). 
It is straightforward to show that the $C_m$ constants are then solutions
of the same linear system shown in Eq.~(\ref{eq-Bound_C}) with
the $B_{im}$ coefficients now defined as in Table~\ref{Tab_B_im2} and:
\begin{equation}
H_{i}=\left\{ \begin{array}{lc}
I^{-}(0,\mu_{i}) & \quad (\mu_{i}<0)\\
0 & \quad (\mu_{i}>0)\end{array}\right.
\end{equation}
\begin{table}[!htbp]
\begin{centering}\begin{tabular}{|c||c|c|}
\hline 
$B_{im}=$&
$m<0$&
$m>0$\tabularnewline
\hline
\hline 
$\mu_{i}<0$&
$T_{im}(0,\mu_{i})$&
$0$\tabularnewline
\hline 
$\mu_{i}>0$&
$0$&
$T_{im}(\tau_{max},\mu_{i})$\tabularnewline
\hline
\end{tabular}\par\end{centering}
\caption{$B_{im}$ coefficients for the one side illumination
boundary conditions in Eq.~(\ref{eq-Bound_C})\label{Tab_B_im2}}
\end{table}


\subsection{Iterative procedure}
At very large optical depths  (e.g. deep inside the cloud or at the core of saturated lines)
the intensity of the radiation field tends to zero. Hence, the simplest way to initiate the 
iterative process is to set  $\mathbf{Q}=\mathbf{R}^{-1}\mathbf{R}'\,\mathbf{y}=0$.
However, this may be far from the real solution, and 
more realistic guesses should be tried. 
In practice, the assumption $\tau \rightarrow \infty$ may be too crude
and one can add the effect of the external radiation
perpendicular to the cloud
that penetrates deepest in the cloud, i.e. attenuated by the smallest
eigenvalue $k_{\pm1}$ (that associated with the radiation field in the $|\mu|\simeq1$ direction).
Thus, we guess a first set of $y_{m}(\tau)$, that we call $y_{m}^{0}(\tau)$, 
from the linear system:
\begin{equation}
\sum_{m=-M}^{-1}R_{lm}(\tau)\, y_{m}^{0}(\tau)+\sum_{m=1}^{M}R_{lm}(\tau)
\, y_{m}^{0}(\tau)=f_{l}^{approx}(\tau) 
\end{equation}
with 
\begin{equation}
f_{0}^{approx}(\tau) = \frac{1}{2}\,I(0,-1)\,\exp[-k_{1}(\tau)\,\tau]+
\frac{1}{2}\,I(\tau_{max},1)\,\exp[k_{-1}(\tau)(\tau_{max}-\tau)]
\label{eq-f_approx}
\end{equation}
Note that only the $l=0$ terms have to be considered.
As noted by Flannery et al. (1980) and Roberge (1983), the presence of dust scattering
implies that $|k_{\pm1}|\neq1$, i.e. 
\textbf{the radiation field attenuation
factor at large depths is not simply $e^{-\tau}$} but dominated by the $e^{-k_1\tau}$ factor. 
This conclusion obviously applies for the present case
with the difference that $k_{\pm1}$ is now depth--dependent and includes line absorption.
This important result can modify the intensity of the
FUV radiation field inside optically thick clouds by orders of magnitude 
depending on the dust grain optical properties. At lower optical depths (e.g., diffuse clouds),
the attenuation factor still contains an important contribution from
additional terms ($k_{\pm2},k_{\pm3},...$).

Now that we have an educated guess for the $y_{m}(\tau)$ variables, we can estimate the
new term in Eq.~(\ref{eq-neweq}) carrying the depth--dependence of the 
gas and dust coefficients, i.e. the
 $\mathbf{Q}=\mathbf{R^{-1}}\,\mathbf{R^{'}}\,\mathbf{y}$ term.
Note that $\mathbf{R}^{-1}\mathbf{R}'$
need to be evaluated only once, so numerical cost is limited. However, special
care should be taken for the $\mathbf{R}'$ derivation. Details of the
$\mathbf{R^{-1}}$ inversion and $\mathbf{R}'$ derivation 
are given in Appendix~\ref{ap_invr}.

We briefly describe the iterative computation of $\mathbf{Q}$:
we start by using $\mathbf{Q^0}=\mathbf{R}^{-1}\mathbf{R}'\,\mathbf{Y^0}$
and then compute a first set of $C_{m}^0$ from the boundary conditions.
With these first $C_{m}^0$ and $\mathbf{Q^0}$ variables we can now use the general
expression Eq.~(\ref{eq-iterate}) to compute a new set of $y_{m}(\tau)$ 
to derive a more refined $\mathbf{Q}$ term, and start this proccess again until
some prescribed level of convergence in $\mathbf{Q}$ is reached.
Thus, if $n$ is the iteration index, $\mathbf{Q}^{(n+1)}$ is computed
from $\mathbf{Q}^{(n+1)}=\mathbf{R}^{-1}\,\mathbf{R}'\,\mathbf{Y}^{(n+1)}$ with:
\begin{equation}
y_{m}^{(n+1)}(\tau)=C_{m}^{(n+1)}\, E_{m}^{-}(\tau)-D_{m}^{-(n)}(\tau) \qquad (m<0)
\end{equation}
\begin{equation}
y_{m}^{(n+1)}(\tau)=C_{m}^{(n+1)}\, E_{m}^{+}+D_{m}^{+(n)}(\tau) \qquad (m>0)
\end{equation}
Those expressions have to be computed at each iteration by  numerical
integration.

\subsection{Mean intensity and FUV photon escape probability\label{sub_Mean-intensity}}

Once we have obtained the full depth and angular description of
the intensity of the radiation field $I(\tau,\mu_i)$ through the
$f_{l}(\tau)$ coefficients, we show here the simple 
form  that $J(\tau)$ takes. The angular average of the specific intensity 
 is defined as:
\begin{equation}
J(\tau)=\frac{1}{2}\int_{-1}^{+1}I(\tau,\mu)\, d\mu
\end{equation}
From the expansion of $I(\tau,\mu)$ we have:
\begin{equation}
\int_{-1}^{+1}I(\tau,\mu)\, d\mu=\sum_{l}(2l+1)\, f_{l}(\tau)\,
\int_{-1}^{+1}P_{l}(\mu)\, d\mu
\end{equation}
where the only no null sum corresponds to $\int_{-1}^{+1}P_{0}(\mu)\, d\mu=2$.
Therefore, as anticipated in Sec.~\ref{subsec-Pl}, the mean intensity
of the radiation field at each wavelength and depth  reduces to  $J(\tau)=f_{0}(\tau)$, that is:
\begin{equation}
J(\tau)=\sum_{m=-M}^{-1}\left(C_{m}\, E_{m}^{-}(\tau)-D_{m}^{-}(\tau)\right)+
\sum_{m=1}^{M}\left(C_{m}\, E_{m}^{+}(\tau)+D_{m}^{+}(\tau)\right)
\end{equation}
where we use the fact that $R_{0m}(\tau)=1$ for all $m$ and $\tau$.
Despite the simplicity of this relation, in many cases of astrophysical
interest (e.g. a two sides illuminated cloud) one needs to distinguish the fraction 
of the radiation field  coming from each side of the cloud. 
In this case, two half sums have to be computed. 
In Appendix~\ref{ap_intensity} we give the analytic formulae to compute the mean
radiation intensity $J^{\pm}(\tau)$ coming from each side.
The resulting $J^{\pm}(\tau)$ values can be  used to evaluate the  escape
probably  of any FUV photon emitted within the cloud, 
e.g. within H$_2$ line cascades. In particular, the probability for a photon emitted
at $\tau$=$\tau'$ (inside the cloud) to reach $\tau$=0 (or  $\tau$=$\tau_{max})$ is given
by the  $J^{-}(\tau')/J^{-}(0)$ (or $J^{+}(\tau')/J^{+}(\tau_{max})$) intensity ratios.
These probabilities can then be further used to determine the H$_2$
level detailed balance.
We also note that in this method the first terms of the intensity
expansion in  Eq.~(\ref{eq-expI})
are directly related to the moments of the radiation field, i.e. $f_{0}(\tau)=J(\tau)$, the
mean intensity;  $f_{1}(\tau)=H(\tau)$, the Eddington flux; and 
$f_{2}(\tau)=3\,K(\tau)-J(\tau)$ where $K(\tau)$ is the $K$--moment.

From the numerical point of view, the  \textit{methodology} described in the previous sections 
has been implemented in the \textit{Meudon PDR code}\footnote{Available at
\textbf{http://aristote.obspm.fr/MIS/}},
a photochemical model of a unidimensional plane--parallel stationary PDR 
(\cite{Jlb93}; \cite{Pet06} and references therein) and will be the 
FUV radiative transfer method used in the code.
In the following sections we  illustrate  several of the  new possibilities
with some relevant astrophysical examples.

\begin{figure*}[th]
    \centering \includegraphics[width=0.55\hsize{}]{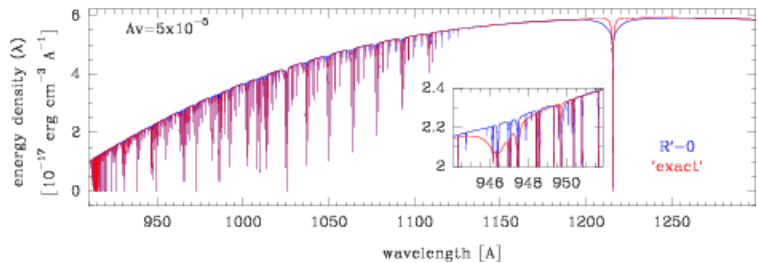}
    \caption{Radiative transfer models for a  cloud with a total extinction 
     of A$_V$=1 and a density of  $n_H$=10$^3$~cm$^{-3}$, illuminated at 
     both sides by the mean ISRF. 
     Part of the resulting FUV spectra ($\sim$912-1300~\AA) close to the cloud surface
     is shown.
     The blue spectra  correspond to a model 
     with R'=0 in Eq.~(\ref{eq-neweq}) (depth dependence neglected), and the red one 
     corresponds to the new "exact" computation.}
    \label{fig:ex-fuv-spectra}
\end{figure*}

\section{Applications: Comparison with previous approaches}\label{sec-ex2}

In this section we compare the main differences of the new \textit{exact} computation
versus the line--plus--continuum approach ($\mathbf{R}'=0$) used by
Le Petit et al. (2006) in the \textit{Meudon PDR code}. Since the previous version
of the code used a single--dust albedo and $g$--asymmetry parameter with no wavelength or
depth dependence, and the extinction curve was not related to the grain properties
used in the model, here we just make
the comparison by assuming $\mathbf{R}'=0$ in the new computation,
and limit ourselves to the uniform dust properties case.  
In the following examples we explicitely include all the H, H$_2$ and CO
electronic absorption lines arising from rotational levels up to $J$=6 (for H$_2$)
and $J$=1 (for CO). The FGK approximation (\cite{Fed79}) is applied for the rest of levels.
Note that the exact method allows one to take into account the overlaps
between H, H$_2$ and CO lines neglected in more crude approaches.

Apart from having a radiative transfer method to consistently solve 
for the dust grain varying populations problem
(Section~\ref{sec-ex1}), the next largest difference between the new computation  compared to
Le Petit et al. (2006) is the effect of line--wing absorption of back--scattered radiation.
At  line core wavelengths, photons penetrating into the cloud are purely absorbed by the gas
(the effective albedo equals~0). 
Due to saturation and opacity broadening, many absorption lines become very wide deep inside
the cloud. As a consequence, the FUV radiation field is more attenuated than in the 
(only) dust continuum transfer case.
At continuum wavelengths free of lines, a fraction of photons coming from the external
 illumination sources can be absorbed by the dust (depending on the exact dust albedo value)  
or be back--scattered (depending on the exact $g$ value) and provide 
an additional  contribution to the radiation field at the cloud surface
(about 10\% of increase for $g=<\cos\theta>\simeq0.6$). 
At line wing wavelengths, where dust and gas opacities are of the same order
(and the effective albedo is in between 0 and the grain albedo),
some of the back--scattered photons 
can again reach the surface of the cloud while another fraction will be absorbed in the
wings. Therefore, as shown by our calculations, line wings are "numerically more challenging".
The fraction of absorbed photons in the line wings depends on the wavelength separation to the line core and on the
transition upper level life time (because it determines the resulting line profile broadening).
To illustrate these differences we  consider a cloud with a constant density
$n_H$=10$^3$~cm$^{-3}$ and a total extinction
depth of $A_V$=1, illuminated at both sides by the 
mean interstellar radiation field (ISRF, $\chi=1$) as defined by \cite{Dra78}.
 These physical conditions
resemble those of a  diffuse cloud such as parts of $\zeta$~Ophiuchi (e.g. \cite{Bla77}).
An uniform grain size distribution similar to that of Mathis et al. (1977) is assumed.
Figure~\ref{fig:ex-fuv-spectra} shows part of the resulting FUV 
spectra ($\sim$912-1300~\AA) close to the cloud surface. 
These spectra clearly show that the effect of H$_2$ line wing absorption of
back-scattered photons is larger in the \textit{exact} computation compared
to the $\mathbf{R}'=0$ approach (\cite{Pet06}).
Note that this is true only for $H_2$ lines.  Atomic hydrogen lines
exhibit the opposite effect, i.e. a decrease of the line wing absorption of back-scattered
photons compared to the $\mathbf{R}'=0$ approach.
Figure~\ref{fig:ex-analysis-diffuse} shows the impact of the same two, \textit{exact}
and $\mathbf{R}'=0$,  computations in the resulting cloud structure
(\textit{left}: H/H$_2$ transition and \textit{right}: 
H$_2$ photodissociation rate). 
In spite of the different  line profiles predicted by each
type of model, the final cloud physical conditions  remain very similar.
Therefore, we conclude that all computations made with the previous
version of the \textit{Meudon PDR code} (\cite{Pet06}), where line transfer was computed (assuming $\mathbf{R}'=0$
and uniform dust properties), are consistent with the present  \textit{exact} 
calculation. 
The larger effect of the H$_2$ line--wing absorptions
in the \textit{exact} calculation increases the attenuation of the
illuminating radiation field, which results in a H/H$_2$ transition layer slightly shifted to
lower extinction depths. 
This general result obviously applies to any FUV radiative
transfer model including gas line absorption compared to (only dust) continuum
models, i.e. the contribution of gas absorption (H$_2$ lines mostly)
decrease the photoionization rate (of neutral carbon particularly) and the photodissociation
rate of species with thresholds close to the Lyman cut.
An  adventage of including gas line absorption is that
predicted spectra can be directly compared with spectral observations
 provided by FUV telescopes.

\begin{figure*}[h]
    \centering \includegraphics[width=6.2cm]{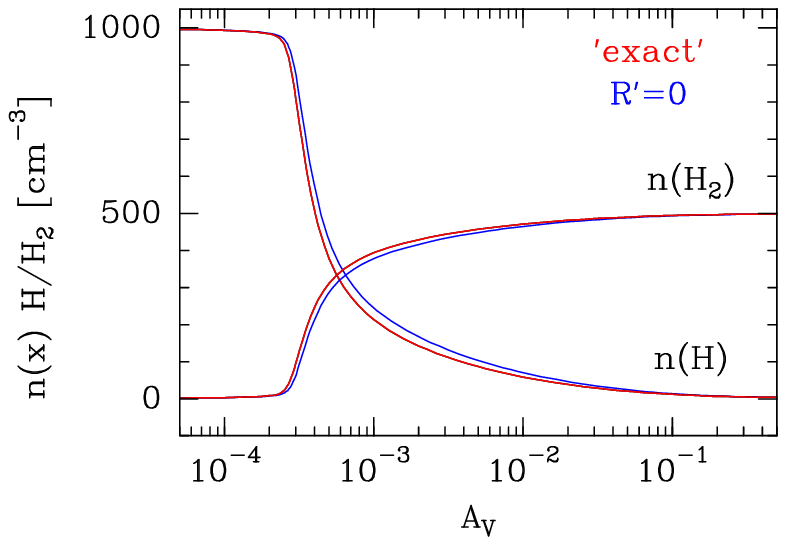}
    \hspace{0.05\hsize{}}
    \centering \includegraphics[width=6.5cm{}]{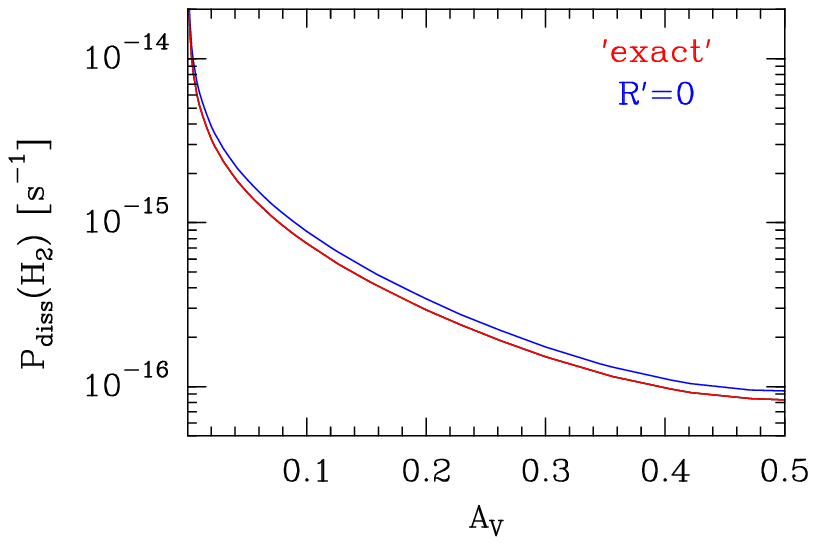}
    \caption{Impact of the new 'exact' radiative transfer computation compared
     to an alternative approach that assumes $R'$=0 (\cite{Pet06}). Grain properties
     are uniform in all the cloud (MRN).
     \textit{Left panel:} H/H$_2$ transition. \textit{Right panel:} H$_2$ photodissociation
     rate as a function of cloud depth.
     A cloud with a density of $n_H$=10$^3$~cm$^{-3}$,
     a total extinction depth of  A$_V$=1  and illuminated at both sides
     by the mean ISRF is considered. These results show that for the case of
     uniform dust grain properties the \textit{error} associated with $R'$=0 assumption
     is small. }
    \label{fig:ex-analysis-diffuse}
\end{figure*}

\clearpage

\section{Applications: Grain growth, varying dust populations}\label{sec-ex1}

\begin{figure*}[b]
    \centering \includegraphics[width=0.4\hsize{}]{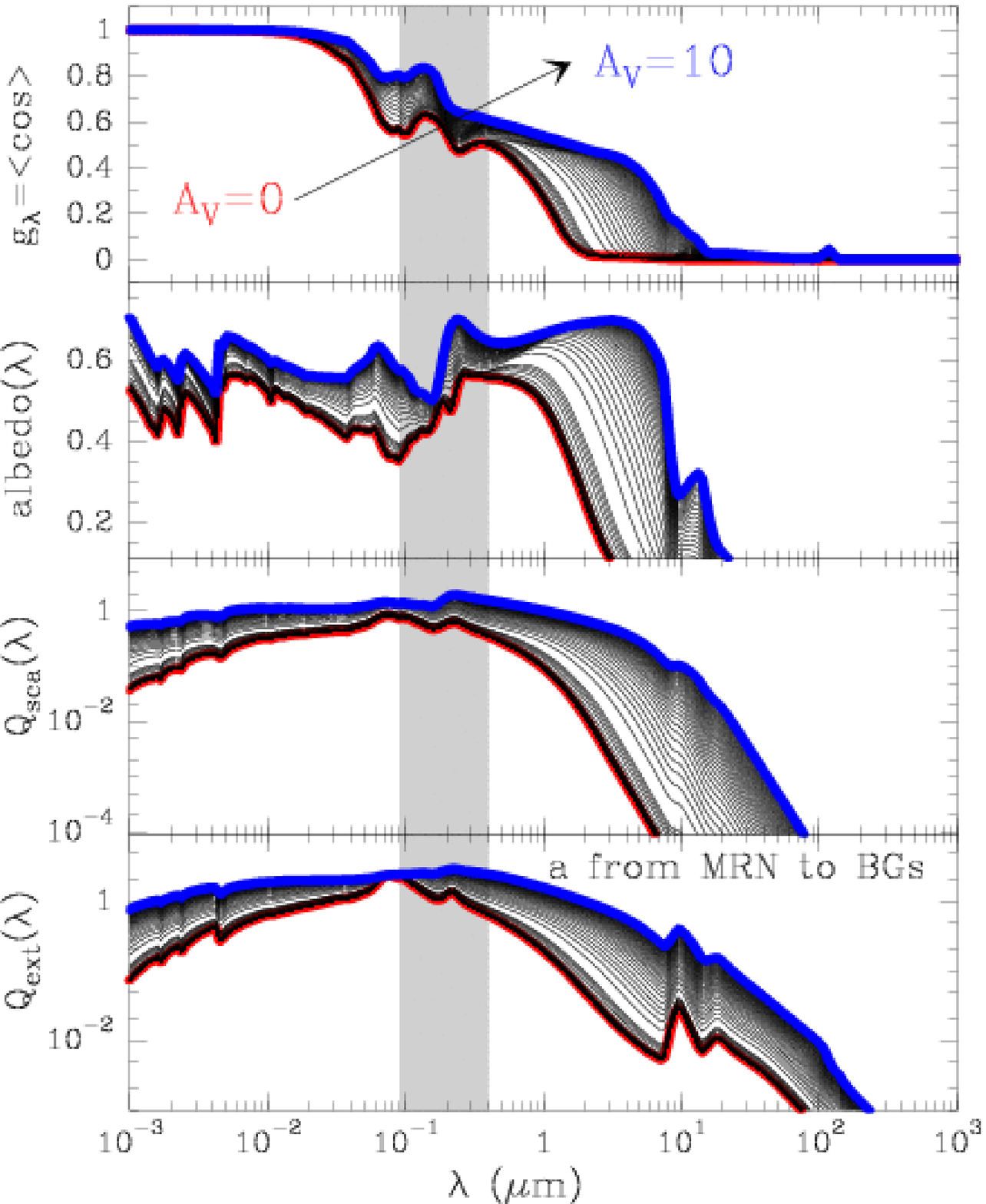}
    \hspace{0.05\hsize{}}
    \centering \includegraphics[width=0.4\hsize{}]{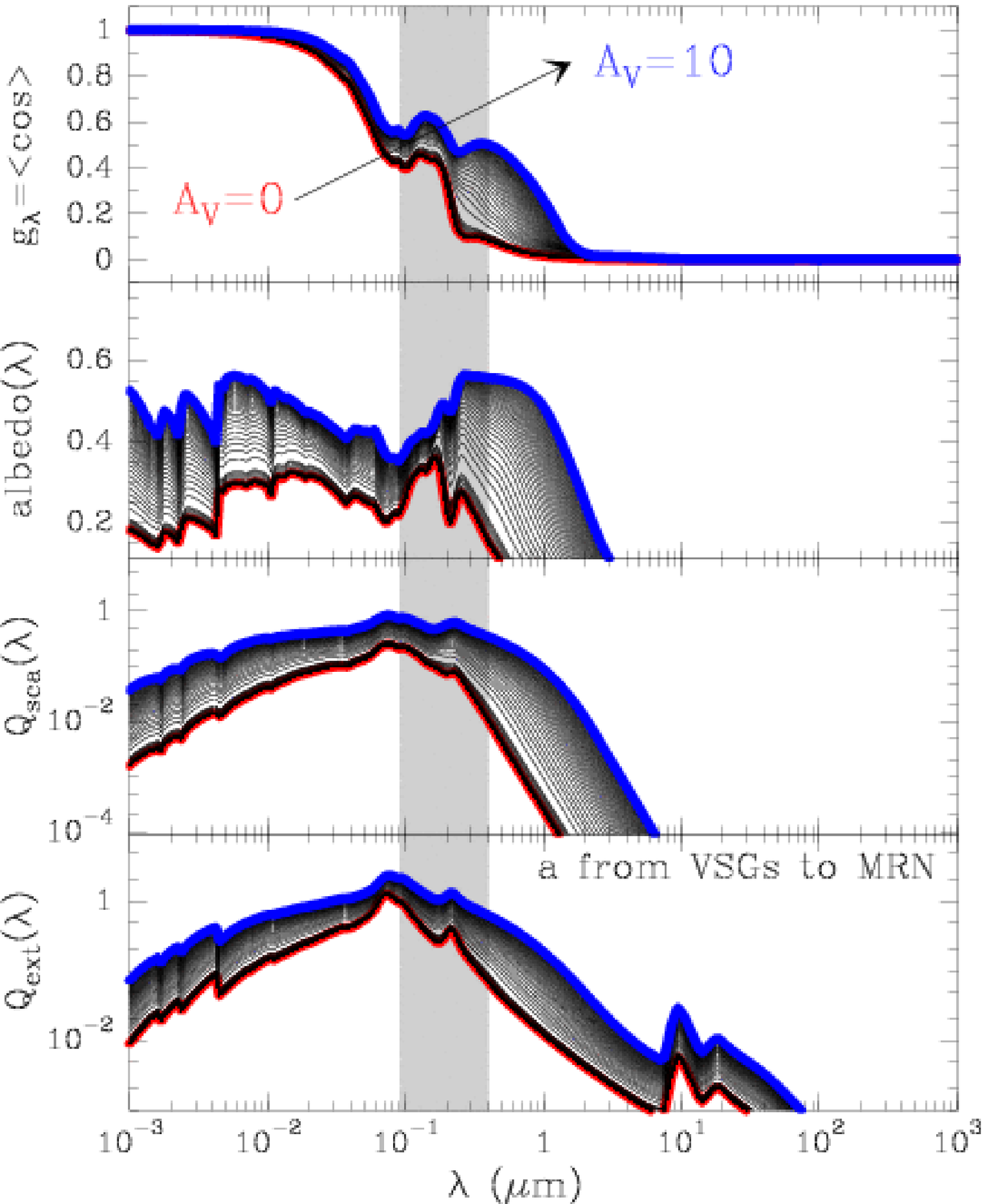}
    \caption{
     Grains mixture optical parameters as a function of
     wavelength and cloud depth (the red curve corresponds to the illuminated
     cloud edge A$_V$=0 and the blue curve to  the center of
     the cloud at A$_V$=10) for the 
    \textit{"MRN to BGs"} (\textit{left}) and  \textit{"VSGs to MRN"} (\textit{right})
    examples respectively.
     The shaded region shows the spectral region taken into
     account in the FUV radiative transfer.}
    \label{fig:ex-props}
\end{figure*}

With the method presented in Sec.~\ref{sec-method}, we can now consistently explore 
the effect of more realistic (non--uniform) dust properties in the FUV penetration into
more embedded objects  e.g., dense molecular clouds or protoplanetary disks.
As a representative example, we present several models  for a dense and strongly 
illuminated cloud (with an ionization parameter of $\chi/n_H=1$~cm$^3$) with grain
 radii varying dust populations. 
From the chemical point of view we only concentrate here on the effects that
the different FUV attenuation depths have on the classical H/H$_2$ and C$^+$/C/CO layered
 structures predicted by  PDR models.
In particular, we consider a cloud with a constant density
$n_H=n(H)+2\,n(H_2)$=10$^5$~cm$^{-3}$ and a total extinction
of $A_V$=20 which is illuminated at both sides by 10$^5$ times the ISRF. 
These physical conditions resemble those of a dense PDR such as the Orion~Bar 
(e.g.~\cite{Tie85}) or a photoevaporating disk around a massive star
(e.g.~Johnstone,  Hollenbach, \& Bally 1998).
At any depth we consider that dust grains follow a size
distribution $dn = n_a\,da$ given by:
\begin{equation}
n_a(\tau) = \sum_i n_{a,i}(\tau)  = \sum_{i} A_i(\tau) \, n_H(\tau)\, a^{-\beta_i}\, da \hspace{1cm} 
a_{i,-}(\tau) < a(\tau) < a_{i,+}(\tau)
\label{eq-mrn-general}
\end{equation}
where a$_{\pm}$ refers to the grain radius distribution lower and upper
limits and $i=1,...,n$ refers to each component of the grain mixture.
In Eq.~(\ref{eq-mrn-general}) we have  explicitly particularized
for the simple power--law case, although more complicated problems may require
other prescriptions of $n_a$ (e.g.~such as those in \cite{Wei01}).
Grain properties were taken from
Laor \& Draine (1993) for silicates and graphite.
With these tabulations we compute the optical parameters of the
grain mixture for each wavelength and cloud depth. In particular,
we compute the $Q_{abs}$, $Q_{sca}$ and
$Q_{ext}$ efficiencies and the grain albedo $Q_{sca}/Q_{ext}$.
We finally use an $g_{\lambda}$--asymmetry factor averaged over the grain distribution 
as (see e.g., \cite{Wol86}):
\begin{equation}
g_{\lambda}(\tau) = \frac{\int_{a_{-}(\tau)}^{a_{+}(\tau)}\, \sum_i \pi\,a^2 \,g_i(a,\lambda,\tau)\,Q_{sca}(a,\lambda,\tau)\,\,n_{a,i}(\tau)\,da}
                       {\int_{a_{-}(\tau)}^{a_{+}(\tau)}\, \sum_i \pi\,a^2\,Q_{sca}(a,\lambda,\tau)\,\,n_{a,i}(\tau)\,da}
\end{equation}

Afterwards, the \textit{extinction curve} 
$A(\lambda,\tau)/A_V(\tau$) and the absolute dust extinction coefficient 
$\alpha^{d}_{\lambda}(\tau)$  are determined at each depth
and used to settle the total line--plus--continuum opacity (as defined in
Eq.~\ref{eq-total-tau}) 
and the effective albedo. 
The dust extinction coefficient (cm$^{-1}$) is given by
$\alpha_d(\lambda,\tau) = n_g\,\pi a^2\, Q_{ext}$,
where $n_g$ is the number of dust grains (per cm$^3$).  Thus, we compute:
\begin{equation}
\alpha_d(\tau) =  \int_{a_-(\tau)}^{a_+(\tau)} \pi\,a^2 \left(  \sum_{i}  Q_{ext}^{i}(a,\tau) 
A_i(\tau)\,n_H(\tau)\,a^{\beta} \right) da
\end{equation} 
The A$_i(\tau)$ grain coefficients are determined at each depth position assuming that the 
gas--to--dust mass ratio has to be constant ($\sim$100) in the whole 
cloud (i.e., the number of grains is reduced
if grain sizes increase). However, in order to keep the grain mixture homogeneous,
the A$_{Sil}$/A$_{Gra}$ ratio is kept fixed. 
Contribution of discrete absorption lines, i.e.
the contribution of $\alpha^{g}_{\lambda}(\tau)$,
is included in similar fashion as described in Le Petit et al. (2006; Sec.~4.3).
The total opacity at each  depth is then given by:
\begin{equation}
d\tau_{\lambda} = \left(1+ \frac{\alpha^{g}}{\alpha^{d}}\right) 
\frac {A(\lambda)}{A_V}\, d\tau_V =
\left(1+ \frac{\alpha_{\lambda}^{g}}{\alpha_{\lambda}^{d}}\right)  
\left(\frac{E_{\lambda-V}} { E_{B-V}}\, \frac{1}{R_V}\, + 1 \right)
\, d\tau_V 
\end{equation} 
where all the variables are depth dependent and where we have assumed
that, in the visible band, the extinction is only produced by dust
and therefore we use
$d\tau_{\lambda} = \frac {A(\lambda)}{A_V}\, d\tau_V = \alpha_{\lambda}^{d}\,ds$
to relate the spatial scale with extinction depth. 
Note that we compute the extinction curve, at each cloud position, directly
from the derived grain properties.

For this "grain growth example" we consider
 that grain radii increase as a function of the cloud depth according to:
\begin{equation}
a_{i,\pm}(\tau) = a_{i,\pm}(0) + 
[a_{i,\pm}(\tau_c) - a_{i,\pm}(0)]\, \left( \frac {\tau}{\tau_c} \right)
^{\gamma_{\pm}} 
\label{eq-growth}
\end{equation}   
where $a_{i,\pm}(0)$ defines the grain radii at the edge of the cloud ($\tau=0$) and
$a_{i,\pm}(\tau_c)$ refers to the grain radii at the center
of the cloud. We chose $\gamma_{\pm}$=2/3. Obviously, this is just an illustrative example
since we do not explicitly solve for the grain nucleation/growth (e.g.~\cite{Sal74})
nor the erosion/sputtering  problem (e.g.~\cite{Bar78}), which depends on the
particular type of source.
The crucial point here is to provide a method to consistently solve for the FUV radiative equation
if, as suggested by observations, the grains size distribution changes toward 
embedded objects (\cite{Moo05}) and/or if  spatial fluctuations 
of the gas to dust ratio do exist along the line of sight (\cite{Pad06}).

\begin{figure*}[t]
    \centering \includegraphics[width=0.25\hsize{},angle=-90]{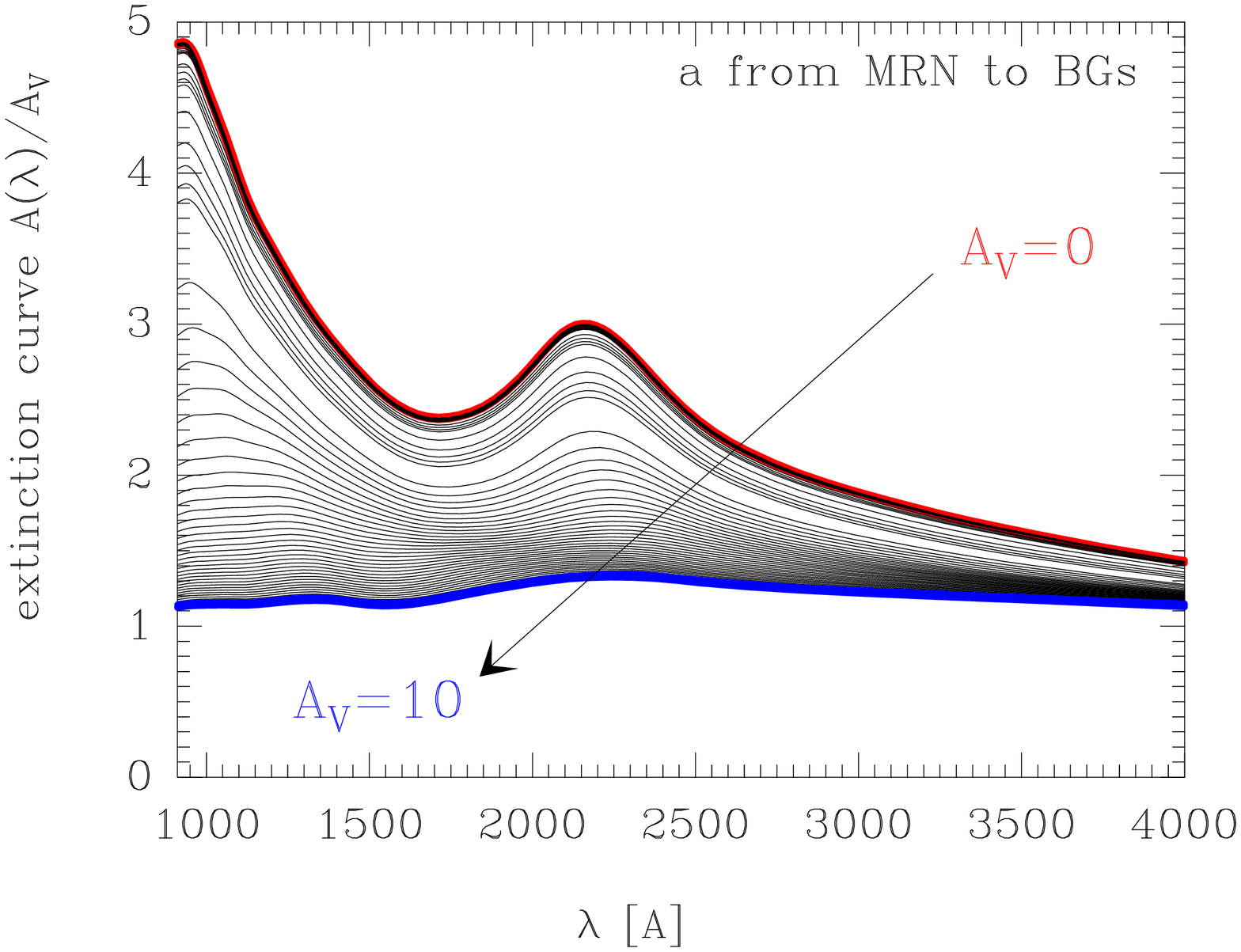}
    \hspace{0.05\hsize{}}
    \centering \includegraphics[width=0.25\hsize{},angle=-90]{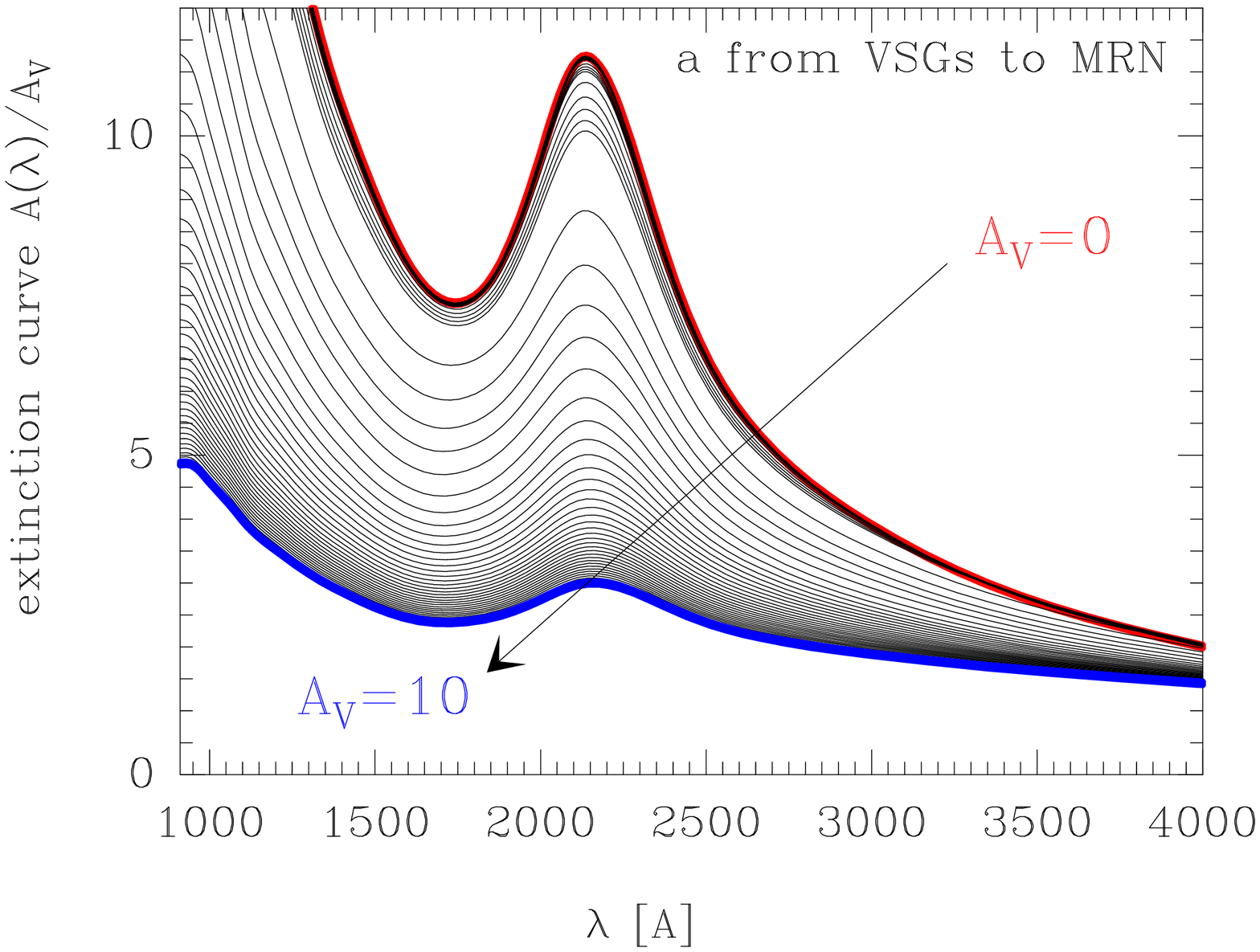}
    \caption{
    Resulting extinction curves as a function of wavelength (only the FUV range 
    is shown) and cloud depth (the red curve corresponds to the
    cloud edge A$_V$=0 and the blue curve to A$_V$=10) for the 
    \textit{"MRN to BGs"} (\textit{left}) and  \textit{"VSGs to MRN"} (\textit{right})
    examples respectively.}
    \label{fig:ex-extinction}
\end{figure*}

\begin{figure*}[h]
    \centering \includegraphics[width=0.25\hsize{},angle=-90]{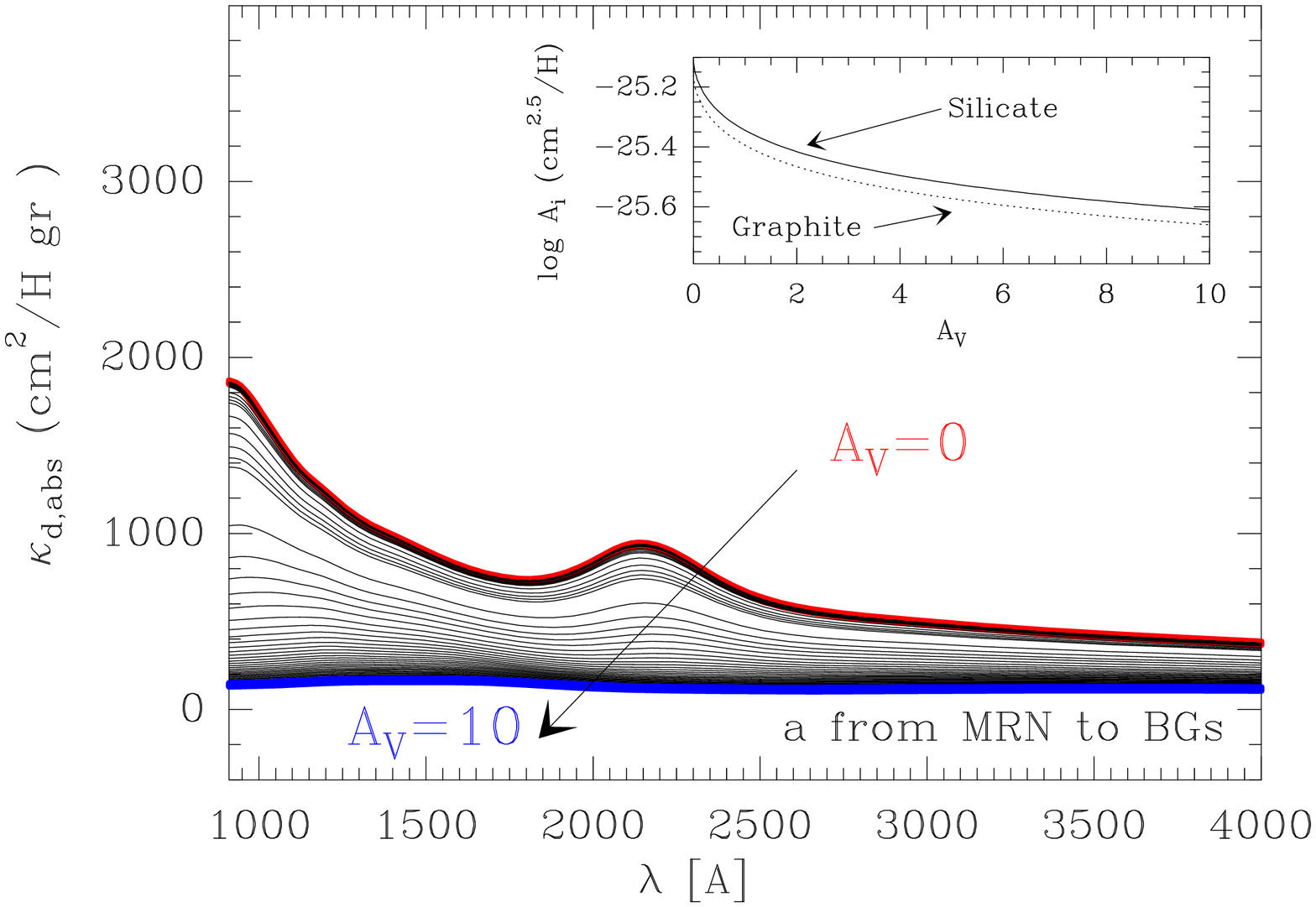}
    \hspace{0.05\hsize{}}
    \centering \includegraphics[width=0.25\hsize{},angle=-90]{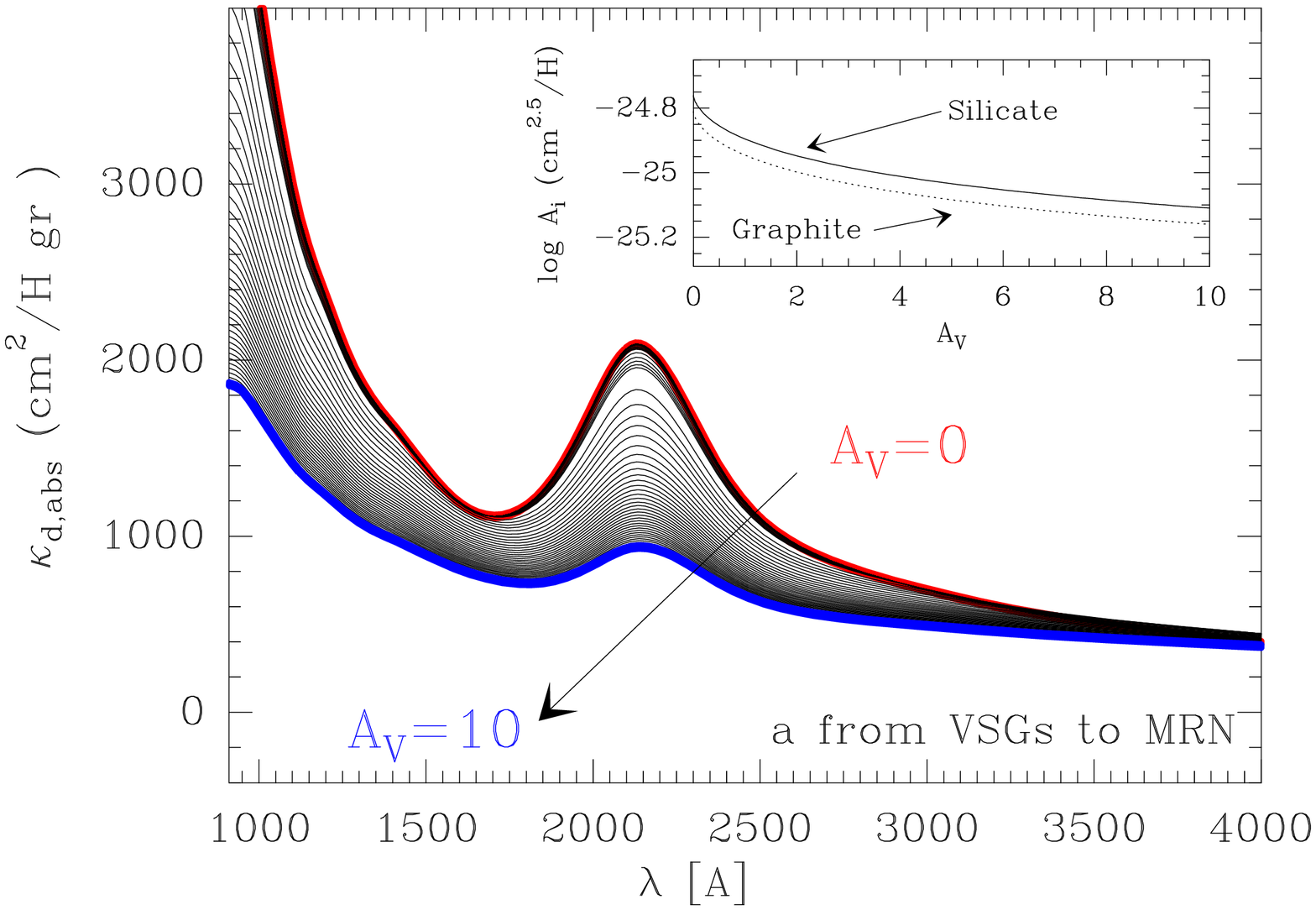}
    \caption{Dust mass absorption coefficients (per gas gr) 
    as a function of wavelength and cloud depth (the red curve corresponds to the
    cloud edge A$_V$=0 and the blue curve to  A$_V$=10) for the 
    \textit{"MRN to BGs"} (\textit{left}) and  \textit{"VSGs to MRN"} (\textit{right})
    examples respectively. The different grain material A$_i$ 
    coefficients required to keep a constant gas--to--dust mass ratio are also shown as a 
    function of $A_V$ in the small insets.}
    \label{fig:ex-massabs}
\end{figure*}

\begin{figure*}[h]
    \centering \includegraphics[width=0.25\hsize{},angle=-90]{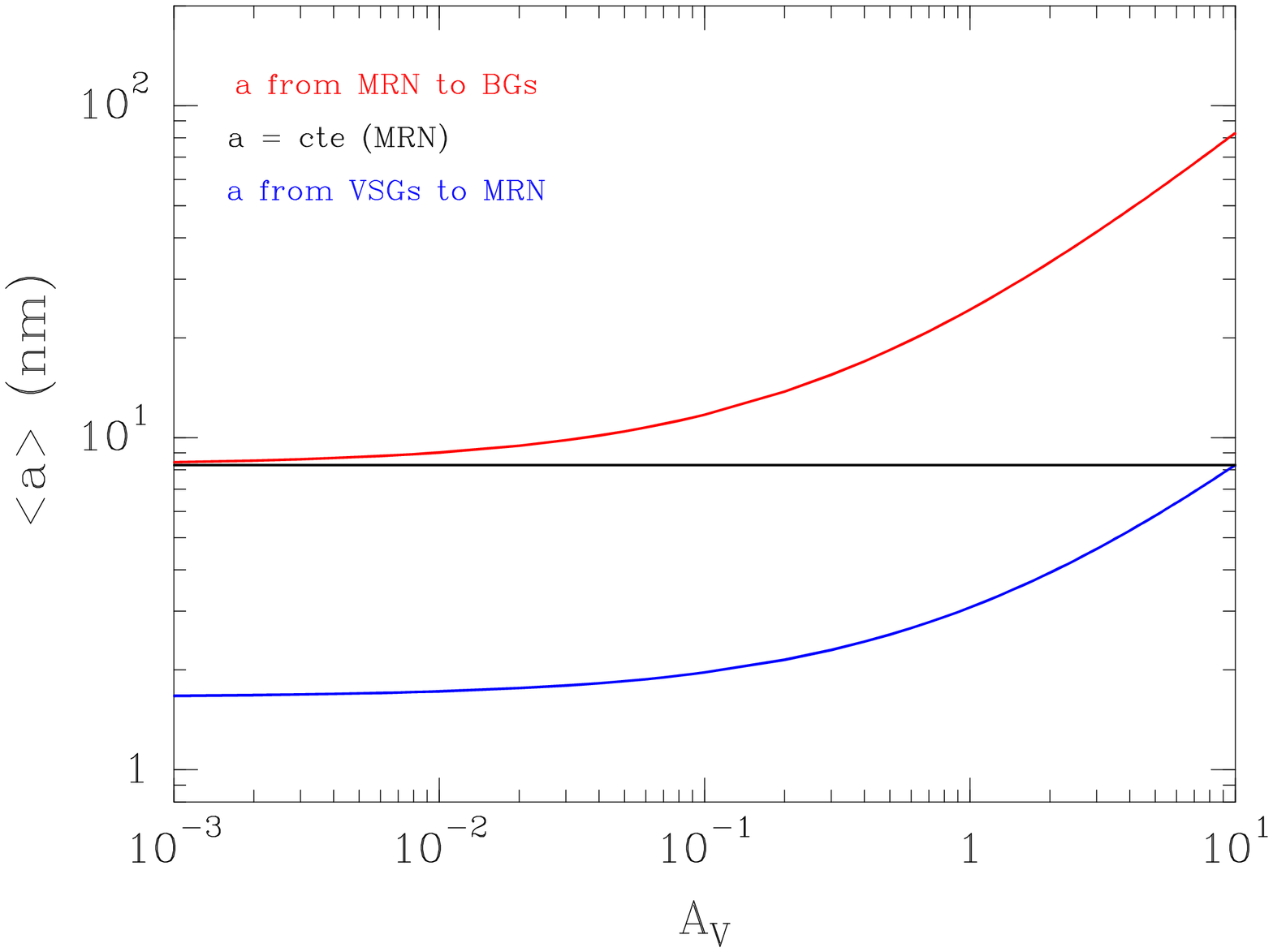}
    \hspace{0.05\hsize{}}
   \centering \includegraphics[width=0.25\hsize{},angle=-90]{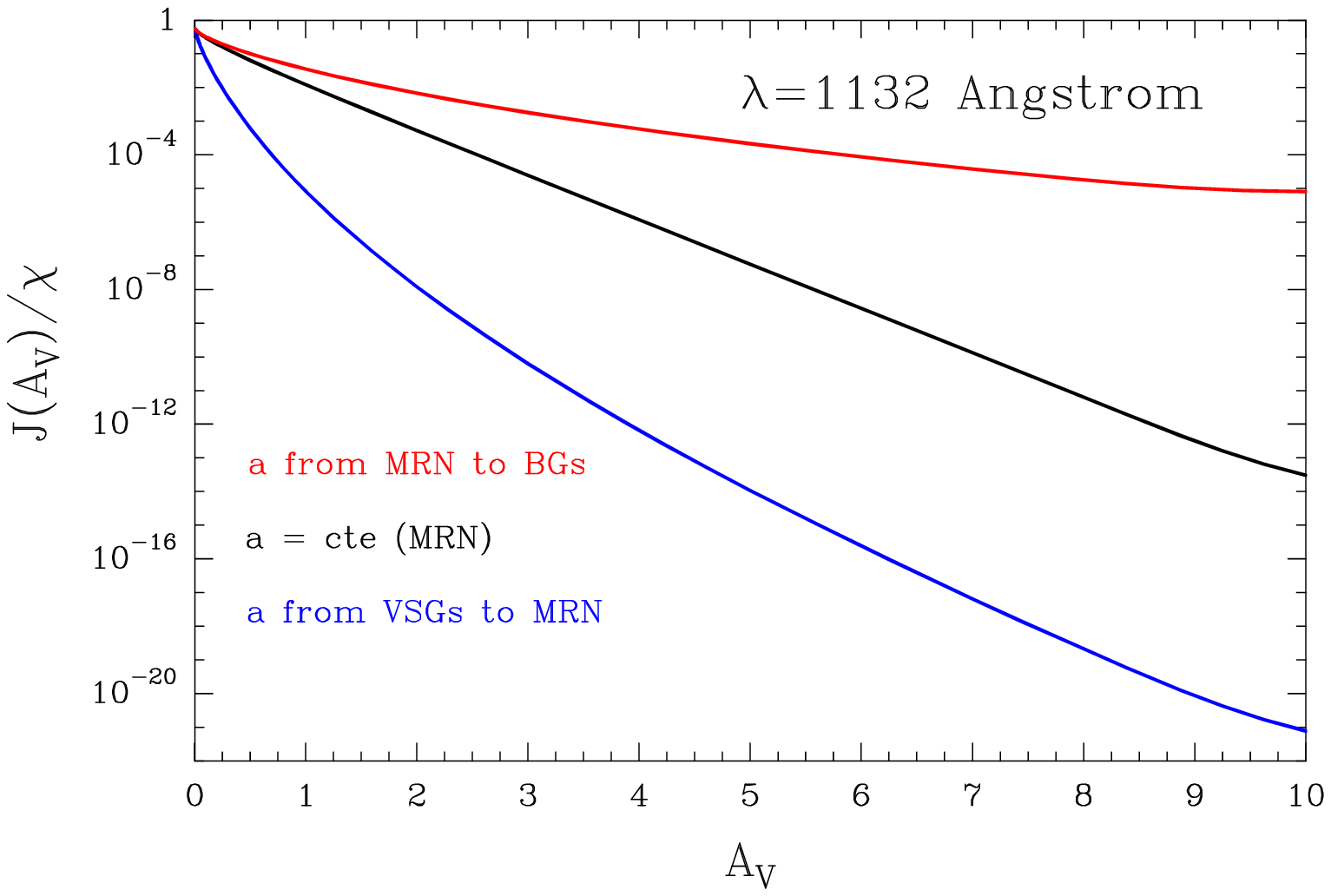}
   \caption{\textit{Left}: Adopted grain averaged radii distribution  
    for the \textit{"MRN to BGs"} (\textit{left}) and  \textit{"VSGs to MRN"} (\textit{right})
    examples respectively.
    \textit{Right}: Resulting mean intensity of the FUV continuum (at $\sim$1132~\AA) as a function of the cloud 
    depth for the three different varying grain populations  discussed in the text.
    The ordinate shows the mean intensity normalized by the illuminating radiation field
    ($\chi$=10$^5$ in Draine's units).}
    \label{fig:ex-radii-inten}
\end{figure*}

In the following, grains follow a power--law distribution of sizes
given by Eq.~(\ref{eq-mrn-general}) with 
$\beta_i$=3.5 at each cloud position.
A mixture of silicates and graphite grains defined by  $A_{Sil}/A_{Gra}=1.1$
and with  $a_-(A_V=0)$=5~nm, $a_+(A_V=0)$=250~nm and
$a_-(A_V=10)$=50~nm, $a_+(A_V=10)$=2500~nm was selected.
Therefore, the grain mixture at A$_V$=0 corresponds
to the size distribution proposed by Mathis, Rumpl \& Nordsieck 
(1977; this size distribution is called MRN hereafter) to fit the
mean galactic extinction curve (see also \cite{Fit90}). 
At A$_V$=10  grains
have grown by a factor 10 and we  call them \textit{Big Grains} (BGs).
In the  second example we only change the 
size distribution to $a_-(A_V=0)$=1~nm, $a_+(A_V=0)$=50~nm and
$a_-(A_V=10)$=5~nm, $a_+(A_V=10)$=250~nm.
Thus, the grain mixture at A$_V$=0 corresponds
to  \textit{very small grains} (VSGs).  At A$_V$=10 grains
have grown by a factor of 5 and  follow a MRN distribution again. 
The third final example considers a uniform grain size distribution (MRN)  
in the whole cloud. The resulting  optical properties, extinction
curves, dust opacities, A$_i$ coefficients and radii distributions for 
these examples are shown in Figs.~\ref{fig:ex-props}, 
\ref{fig:ex-extinction},
\ref{fig:ex-massabs} and \ref{fig:ex-radii-inten} (\textit{left panel}), respectively.

\begin{figure*}[h]
    \centering \includegraphics[width=0.65\hsize{}]{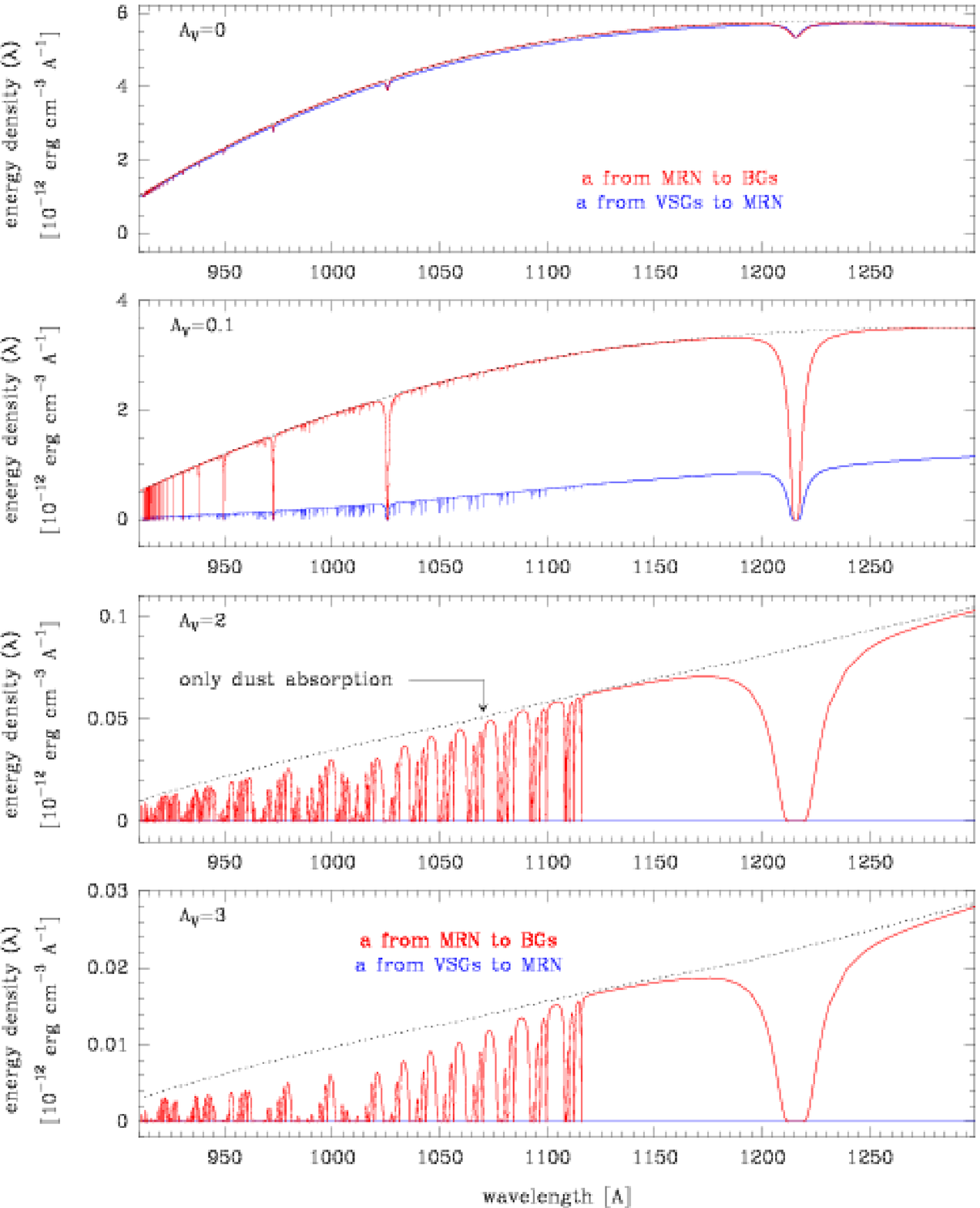}
    \caption{Radiative transfer models for a  cloud with a total extinction 
     of A$_V$=20 and a density of  $n_H$=10$^5$~cm$^{-3}$, illuminated at 
     both sides by 10$^5$ times the mean ISRF. 
     Part of the resulting FUV spectra ($\sim$912-1300~\AA) at
     different extinction depths: A$_V$=0  (cloud surface), A$_V$=0.1,
     A$_V$=2 and A$_V$=3 are shown in each box. 
     In each panel, the red (blue) curve corresponds
     to the  \textit{"MRN to BGs"} (\textit{"VSGs to MRN"}) example.}
    \label{fig:ex-fuv-spectra-bar}
\end{figure*}

Some time ago, Sandell \& Mattila (1975) emphasized that the albedo and anisotropy
of dust grain scattering have important effects on photodissociation rates
for ISM molecules.
The  present computation of the  FUV radiation field (continuum+lines) at each
cloud position (see Fig.~\ref{fig:ex-fuv-spectra-bar} for the resulting FUV spectra
at different A$_V$) allows an explicit integration of 
consistent  C  photoionization  rates together with H$_2$ and  CO
photodissociation rates.
Once the FUV radiation field has been determined and the photo rates
calculated, steady-state chemical abundances 
are computed  for a given network of chemical reactions. 
Finally, we compute the thermal 
structure of the cloud by solving the  balance between the most 
important gas heating and cooling  processes 
(\cite{Jlb93}; \cite{Pet06} and references therein).

\begin{figure*}[ht]
    \centering \includegraphics[width=0.95\hsize{}]{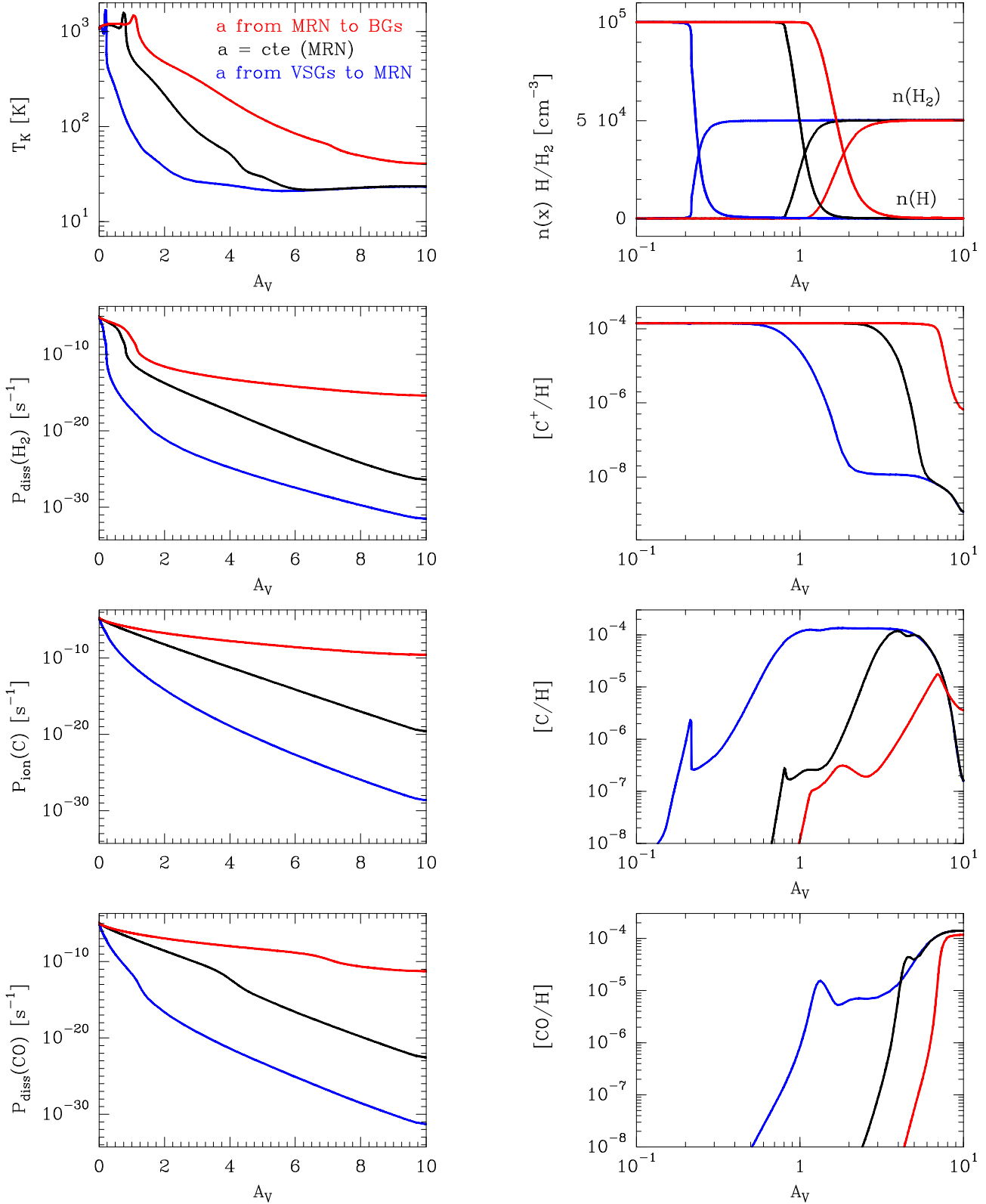}
    \caption{Impact of the different FUV radiative transfer models on the
     kinetic temperature, H$_2$ photodissociation rate, C photoionization rate 
     and CO  photodissociation rate
     (\textit{left column}), H/H$_2$ transition and C$^+$/C/CO column densities 
     (\textit{right column}). A cloud with a density of $n_H$=10$^5$~cm$^{-3}$,
     a total extinction depth of  A$_V$=20  and illuminated at both sides
     by 10$^5$ times the mean ISRF is considered. Although not clearly seen
     in these boxes, all physical parameters show an horizontal tangent at Av = 10, 
     consistent with their null variation  with respect to the depth position at half cloud
     (as expected for a symmetrically illuminated cloud).}
    \label{fig:ex-analysis}
\end{figure*}

Depending on the grain properties these examples
show FUV radiation fields that change by orders of magnitude at large A$_V$
(Fig.~\ref{fig:ex-radii-inten} \textit{right panel}). 
Note that the mean radiation intensity  at the cloud surface $J(0)$
cannot be larger than the illuminating radiation field itself, i.e., $J(0)/\chi<1$.
The exact ratio depends on the particular dust scattering properties ($\sim$0.53-0.54 for
these models of optically thick clouds). 
The influence of the different grain
distributions in the  attenuation of FUV radiation is obvious,
the FUV penetration depth is larger when  dust scattering is more 
efficient, i.e., when  grain albedo and  scattering anisotropy increase
(as dust grains grow toward bigger grains).
Note  that the only difference between models is the change of  grain
size distributions across the cloud. 
Therefore, the assumption of uniform dust properties and averaged extinction curves
can be one of the crudest approximations made to determine the 
resulting cloud physical and chemical state.  
Figure~\ref{fig:ex-analysis} shows the impact of the different
grain growth curves on the resulting cloud structure: 
kinetic temperature, H$_2$ photodissociation rate, C photoionization rate and CO
photodissociation rates (\textit{left column}), 
H/H$_2$ transition,  and C$^+$/C/CO abundances (\textit{right column}).  
The different intensities of the FUV radiation field for each dust population 
result in very different photoionization and photodissociation rates
which ultimately determine the prevailing chemistry.
This conclusion qualitatively agrees with earlier calculations for ISM diffuse clouds
(\cite{Rob81}) and should be extended to more embedded objects where there
are observational evidences (e.g. \cite{Moo05})
of flatter extinction curves (consistent with grain growth). 
The H/H$_2$ and C$^+$/C layered structures in our models are different  
even in similar sources (same density and illumination) if grain properties significantly
disagree, or if dust grains vary along the observed region. 
Different ionization fractions, molecular ions enhancements, 
and C$^+$/C/CO abundances should thus be observed.
In particular,  photochemistry can still be important
at large~A$_V$  if anisotropic scattering of the illuminating radiation is efficient 
(e.g., "MRN to BGs" model). In this case, CO photodissociation and carbon ionization still dominate
the CO destruction and C$^+$ formation respectively deeper inside the cloud.
As a result, the predicted abundance of neutral and ionized carbon at A$_V$=10 is enhanced compared 
to standard MRN dust models
(see Fig.~\ref{fig:ex-analysis}).

Secondly, the intensity of the FUV radiation field also  determines much of
the thermal structure
of the cloud through the efficiency of the grain photoelectric effect,
the dominant heating mechanism (e.g. Draine 1978).  
Since FUV radiation penetrates deepest when dust grains
are bigger, the photoelectric heating rate is kept high
deeper inside the cloud. Thus, a larger fraction of the gas is maintained warm 
at large extinction depths. Warmer temperatures also affect the rates of chemical reactions
with activation energy barriers.
For the smallest dust grains, FUV attenuation is so high that photoelectric heating
soon becomes inefficient and the gas is colder at large extinctions depths.
Note that since grain ionization is very large in the surface of the cloud 
(due to the high illumination in the selected example), the maximum efficiency of the 
photoelectric effect, i.e. the maximum temperature, is reached deeper inside the cloud 
where the grain ionization has decreased.
The general effects described here must play a significant role in illuminated sources
where grain growth takes place, specially in protoplanetary disks,
circumstellar envelopes around evolved stars and dense molecular clouds near H\,{\sc ii} regions.
In these cases, the FUV penetration depth is increased if dust grains
evolve toward bigger grains, leading to larger photochemically active regions.

Conversely, molecular species such as CO will be more abundant 
in irradiated regions where the smallest grains dominate the extinction efficiency.
Figure~\ref{fig:ex-analysis-diffuse} shows the effects 
of grain growth in a diffuse cloud (A$_V$=1), with a density of $n_H$=10$^3$~cm$^{-3}$, and
illuminated by the mean ISRF. Although the resulting variations are not so large compared
to optically thick clouds, the different photoionization and photodissociation rates
also translate into different atomic and molecular abundances. 

In particular, the
C$^+$/C and C$^+$/CO abundance ratios change up to a factor $\sim$10 depending on the
assumed grain properties. Note that for optically thin clouds, the mean intensity 
at one surface can have a significant contribution from the other side illumination 
(that increases with the scattering efficiency). As an example, the mean intensity
at A$_V$=0 in the "MRN to BGs" grain model ($J(0)/\chi$$\simeq$0.63; red curves in 
Fig.~\ref{fig:ex-analysis-diffuse}) is a factor $\sim$$20\%$ larger than in the "VSGs to MRN" 
model (blue curves). This effect slightly modifies the dissociation and 
ionization rates at the cloud surface.

In summary, as gas photodissociation and heating 
determine much of the  chemistry in FUV irradiated gas, 
the resulting source structure is
severely altered by the assumed (or observed) grain properties.
Therefore, understanding  dust properties and
grain variations in individual sources is a crucial step to determine
the source physical and chemical state.

\clearpage

\begin{figure*}[ht]
    \centering \includegraphics[width=0.95\hsize{}]{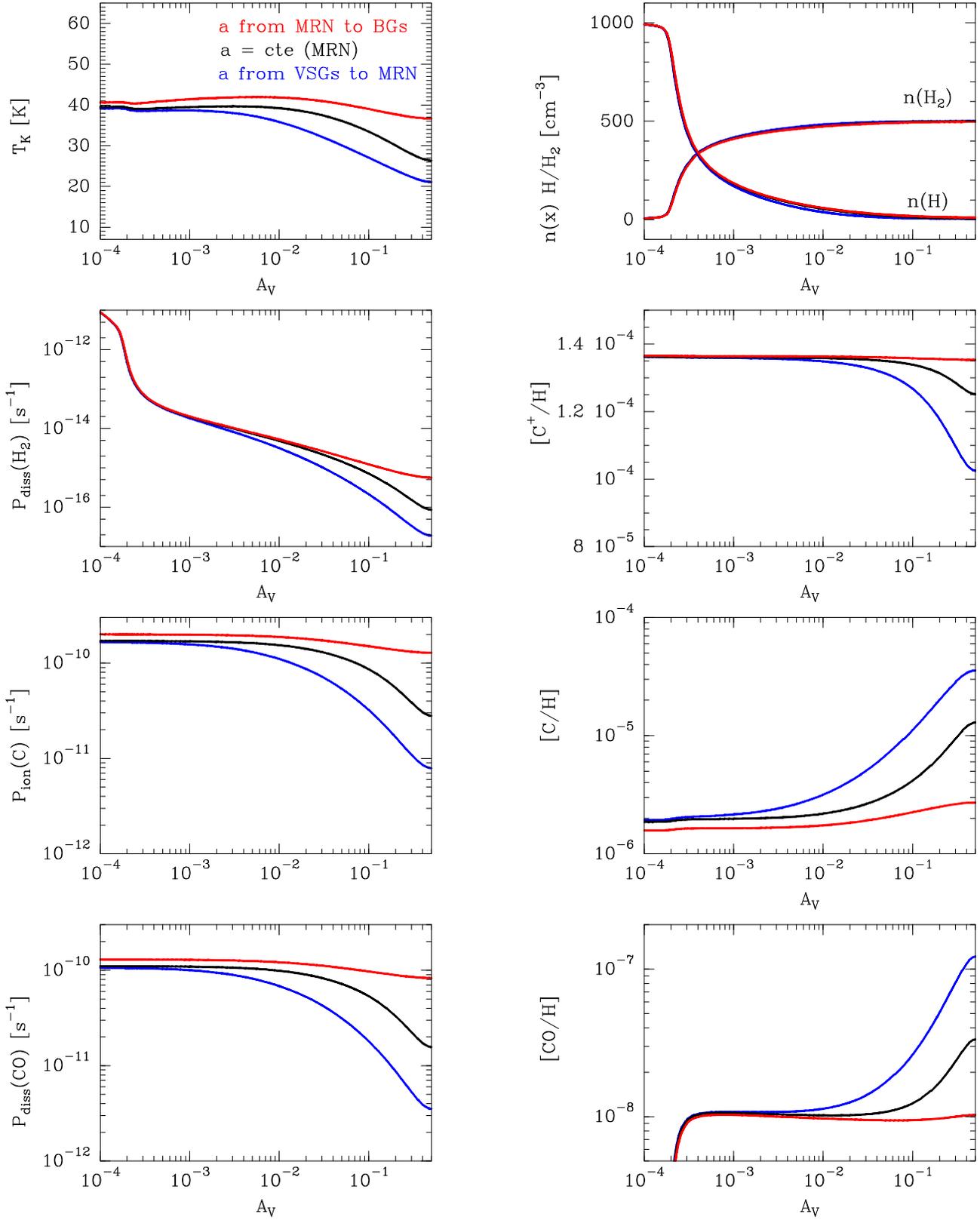}
    \caption{Same as Figure~\ref{fig:ex-analysis} for a cloud with 
     $n_H$=10$^3$~cm$^{-3}$, a total extinction depth of A$_V$=1 and illuminated at both sides
     by the the mean ISRF. Dust grains grow according to Eq.~(\ref{eq-growth}) with
     A$_{V}^{c}$=1.086~$\tau_{V}^{c}$=0.5.}
    \label{fig:ex-analysis-diffuse}
\end{figure*}

\clearpage

\section{Summary and conclusions}

\begin{enumerate}

\item An  extension of the spherical harmonics 
method to solve for the radiative transfer equation with depth
dependent coefficients in plane--parallel geometry has been presented.
The  method can be used to solve for the FUV radiation
field  in externally or internally illuminated clouds, 
taking into account gas absorption and coherent, nonconservative
and anisotropic scattering by dust grains. Our extended formulation thus
allows to consistently include $(i)$ gas lines and
$(ii)$ varying dust populations.

\item We have shown that the penetration of FUV radiation  
is heavily influenced by dust properties.
According to the dust ISM and CSM life--cycle, such properties likely change from
source to source but also they change within the same object.
The FUV penetration depth rises for increasing dust albedo and 
anisotropy of the scattered radiation
when grains grow at large A$_V$ (as suggested observationally).
Therefore, the modeled physical and chemical state of illuminated molecular clouds, 
protoplanetary disks
or entire galaxies can be altered by large factors if a more realistic treatment of the interaction
between radiation and matter is considered.

\item  The new formulation has been implemented in the \textit{Meudon PDR code}
and thus it will be publicly available.
Particular examples where only the dust populations are changed
show intensities of the FUV radiation field that differ by orders of magnitude 
at large~A$_V$. Therefore, the resulting photochemical and thermal structures of molecular clouds 
can be very different 
depending on the assumed grain properties and growth.

\end{enumerate}

\begin{acknowledgements}
 We warmly thank G. Pineau~Des~For\^ets, P. Hily-Blant, 
F. Le Petit, M. Gerin and M. Walmsley for useful discussions or suggestions. 
We thank our referees for useful comments.
JRG was supported by a \textit{Marie Curie Intra-European Individual
Fellowship} within the 6th European Community Framework Programme,
contract MEIF-CT-2005-515340.  
\end{acknowledgements}


\clearpage

\begin{appendix}

\section{Inclusion of embedded sources of emission ($S^*\neq0$)\label{ap_embedded}}

In this appendix we give the recipe to include the true emission by "embedded sources
of photons" in the method described
in section~\ref{sec-method}. In this case the source function 
 includes the scattering of
photons by dust grains plus a non null 
$S^{*}_{\lambda}=\frac{j_{\lambda}}{\alpha_{\lambda}+\sigma_{\lambda}}$ term
(see Eq.~\ref{eq-etr2} and Fig.~\ref{fig:geo}), where $j_{\lambda}(s)$ is the emission coefficient 
(line or continuum) of any source of internal radiation. 

Firstly, the angular dependence of $S^{*}(\tau,\mu)$ has to be also expanded in a truncated series 
of Legendre polynomials $P_{l}(\mu)$ as:
\begin{equation}
S^*(\tau,\mu)=\sum_{l=0}^{L}(2l+1)\, s_{l}(\tau)\, P_{l}(\mu)
\label{eq-expS}
\end{equation}
where the dependence with $\lambda$ is omitted.
The inclusion of Eqs.~(\ref{eq-expS}) into the transfer equation (\ref{eq-etr2})
leads to an additional term in the set
of coupled, linear, first order differential equations in the 
unknown $f_l(\tau)$ coefficients:
\begin{equation}
lf'_{l-1}(\tau)+(l+1)f'_{l+1}(\tau)=(2l+1)\,[1-\omega(\tau)\sigma_{l}(\tau)]
\, f_{l}(\tau)- (2l+1)\,s_l(\tau)
\label{eq-system-embedded}
\end{equation}
Therefore, the system of equations (\ref{eq-system-embedded}) is now non-homogeneous and can
be written as:
\begin{equation}
\mathbf{f'}(\tau)=\mathbf{A^{-1}\mathnormal{(\tau)\,}f}(\tau)+\mathbf{A^{-1}\mathnormal{(\tau)\,}g}(\tau)
\label{eq-system2-embedded}
\end{equation}
where $g_{l}(\tau)=-s_{l}(\tau) /(1-\omega(\tau)\,\sigma_{l}(\tau))$.
Although the method can be easily used for anisotropic source functions,
in most practical applications, the embedded sources of photons emit isotropically 
and therefore the terms in the expansion of $S^*(\tau)$ in Eq.~(\ref{eq-expS}) 
reduce to $s_{l}(\tau)=S^*(\tau)\,\delta_{l0}$, and thus,
$g_{l}(\tau)$ reduces to $g_{0}(\tau)=-S^*(\tau)/(1-\omega(\tau))$ with 
$g_{l}(\tau)=0$ if $l\neq0$. 
Using the same set of auxiliary variables  $\mathbf{y}(\tau)=\mathbf{R^{-1}}(\tau)\,\mathbf{f}(\tau)$,
Eq.~(\ref{eq-neweq}) is now written as:
\begin{equation}
\mathbf{y'}=\mathbf{K}\,\mathbf{y}-\mathbf{R}^{-1}\,\mathbf{R}'\,\mathbf{y}
+\mathbf{K}\,\mathbf{R}^{-1}\,\mathbf{g}
\label{eq-neweq-embedded}
\end{equation}
Note that by inserting $\mathbf{R}\mathbf{R}^{-1}$ between $\mathbf{A}^{-1}$
and $\mathbf{g}$, we have simplified $\mathbf{R}^{-1}\,\mathbf{A}^{-1}\,\mathbf{g}$
as $\mathbf{K}\,\mathbf{R}^{-1}\,\mathbf{g}$.
This result is particularly useful\footnote{This is true whatever the isotropy 
properties of the source function are, and not only for the isotropic case.},
since it avoids computing $\mathbf{A}^{-1}$ completely.
Hence, the last matrix product,  
$\,\widetilde{\mathbf{Q}}=\mathbf{K}\,\mathbf{R}^{-1}\,\mathbf{g}$,
 makes the system non--homogeneous: 
\begin{equation}
y'_{m}=k_{m}(\tau)\, y_{m} + [\mathbf{L\,y}]_m(\tau)+\widetilde{q}_{m}(\tau)
\label{eq-new-eq-embedded}
\end{equation} 
Eq.~(\ref{eq-new-eq-embedded}) can also be solved with the iterative scheme 
described in section~\ref{sec-method} by including the additional $\widetilde{q}_{m}$ term, i.e.,
\begin{equation}
y_{m}^{(n+1)}(\tau)=e^{\int_{\tau_{m}}^{\tau}\,k_{m}(t)\,dt} 
\left[ C_{m}^{(n+1)} + \int_{\tau_{m}}^{\tau}\, e^{-\int_{\tau_{m}}^{t}\,k_{m}(t')\,dt'} \,
\left( [\mathbf{L\,y}^{(n)}]_m(t) + \widetilde{q}_{m}(t) \right) dt \right]
\label{eq-iterate-embedded}
\end{equation}
It is straightforward to show that the $f_l(\tau)$ terms in the Legendre expansion of
the radiation field $I(\tau,\mu)$ are still given by Eq.~(\ref{eq-final_f}). The only change
compared to the $S^*$=0 case is that the $q_{m}(\tau)$ variables in the
$D_{m}^{-}(\tau)$ and $D_{m}^{+}(\tau)$ integrals (Eqs.~\ref{eq-D-} and \ref{eq-D+}) have to be substituted
by $q_{m}(\tau)-\widetilde{q}_{m}(\tau)$, that is:
\begin{equation}
D_{m}^{-}(\tau)=\int_{0}^{\tau}\frac{E_{m}^{-}(\tau)}{E_{m}^{-}(t)}\,\left(q_{m}(t)-
\widetilde{q}_{m}(t)\right)\, dt\qquad (m<0)
\end{equation}
\begin{equation}
D_{m}^{+}(\tau)=\int_{\tau}^{\tau_{max}}\frac{E_{m}^{+}(\tau)}{E_{m}^{+}(t)}\,
\left(q_{m}(t)-\widetilde{q}_{m}(t)\right)\, dt\qquad (m>0)
\end{equation}
The iterative procedure can now  be initiated taking into account that at large optical depths
the intensity of the radiation field is isotropic and tends to the ratio of the true emission 
to the true absorption:
\begin{equation}
I(\tau \rightarrow \infty) \simeq \frac{S^*(\tau)}{1-\omega(\tau)}\delta_{l0}
\end{equation}
In practice, the assumption $\tau \rightarrow \infty$ may be too crude.
We have computed that by adding the effect of the external radiation
that penetrates deepest into the cloud, the iterative scheme is more robust.
Therefore, the first set of $y_{m}(\tau)$ variables in the iterative procedure, $y_{m}^{0}(\tau)$, 
are computed from the linear system:
\begin{equation}
\sum_{m=-M}^{+M}R_{lm}(\tau)\, y_{m}^{0}(\tau)
=f_{l}^{approx}(\tau) =
\frac {s_{0}(\tau)}{1-\omega(\tau)}+\frac{1}{2}I(0,-1)\,\exp[-k_{1}(\tau)\,\tau]+
\frac{1}{2}I(\tau_{max},1)\,\exp[k_{-1}(\tau)(\tau_{max}-\tau)]
\end{equation}
where only the $l=0$ terms are considered.

We have successfully applied the above method by associating $S^*$ to thermal emission of dust.
These kind of computations are useful if the radiative transfer calculation is extended to the 
IR domain ($\lambda>1~\mu$m), where scattering of IR photons by
dust grains is still significative.
In the FUV domain, $S^*$ can represent any source of internal illumination.
In a future paper we plan to include "secondary" line photons in the embedded 
source function. This \textit{line FUV radiation field} arises from
the H$_2$ radiative de-excitations that follow the H$_2$ excitation by collisions 
with electrons and cosmic rays (\cite{Pra83}) and is generally poorly treated. However, 
 it constributes to molecular photodissociation deep inside
molecular clouds where the \textit{continuum FUV radiation field} has been attenuated.



\section{Numerical solution and error limits\label{ap_error}}

In Sec.~\ref{sub-sec-numsol} we turn the system of differential equations~(\ref{eq-new-eq})
into and integral problem (Eqs.~\ref{eq-integral-form}) that we solve numerically
through an iterative scheme (Eqs.~\ref{eq-iterate}). In this appendix we provide
a bound on the error associated with this procedure and we verify that the derived
solution satisfies the original system of equations (\ref{eq-system2}).

Given a numerical approximation $y_{m}^{(n)}$ to the true solution $y_m$, we
investigate if our iterative proccess converges for all $\lambda$ and $A_V$ of the
wavelength and cloud depth grids.
Thus, we compute:
\begin{equation}
y_{m}^{(n+1)}(\tau)=e^{a_m(\tau)} 
\left[ C_{m}^{(n+1)} + \int_{\tau_{m}}^{\tau}\, e^{-a_m(t)} \,
 [\mathbf{L\,y}^{(n)}]_m(t)\, dt \right]
\hspace{.8cm}$with$\hspace{.8cm}
 \left[\mathbf{Ly}^{(n)}\right]_{m}(t)=\sum_{i=-M}^{+M}L_{mi}(t)\, y_{i}^{(n)}(t)
\label{eq-app-iterate}
\end{equation}
and write the error in step $n+1$ as $\Delta_{m}^{(n+1)}=y_{m}-y_{m}^{(n+1)}$
where $m=\pm 1,...,\pm M$. Note that this is the difference between the true $y_m$
(unknown) and our numerical approximation at step $n+1$. Since the above equations
are linear, $\Delta_{m}^{(n+1)}$ reduces to:
\begin{equation}
\Delta_{m}^{(n+1)}(\tau)\simeq e^{a_{m}(\tau)}\,\sum_{i=-M}^{+M}\,
\left(\int_{\tau_{m}}^{\tau}e^{-a_{m}(t)}\, L_{mi}(t)\,\Delta_{i}^{(n)}(t)\, dt\right)
\label{eq_error}
\end{equation}
because the boundary conditions term ($C_m-C_{m}^{(n+1)}$) is small and damped almost
everywhere by the exponential term (as shown numerically).
If we now define $\Delta_{i}^{(n),MAX}=\max_{t}|\Delta_{i}^{(n)}(t)|$,
the maximum error at iteration step $n$ in the Legendre expansion of order $i$
($i=0,1,...,L$) at any depth position, then:
\begin{equation}
\Delta_{m}^{(n+1),MAX}<\sum_{i=-M}^{+M}\,\Delta_{i}^{(n),MAX}\,
\left(\int_{\tau_{m}}^{\tau}e^{\left[a_{m}(\tau)-a_{m}(t)\right]}\, L_{mi}(t)\, dt\right)
\end{equation}
By taking the maximum error at iteration step $n$ at any depth position and at any
Legendre order, 
$\Delta^{(n),MAX}=\max_{m}\Delta_{m}^{(n),MAX}$,
 we arrive to a severe upper limit to the error between the true
solution and the numerical approximation at step $n+1$:
\begin{equation}
\Delta^{(n+1),MAX}<\Delta^{(n),MAX}\,\sum_{i=-M}^{+M}\,
\left(\int_{\tau_{m}}^{\tau}e^{\left[a_{m}(\tau)-a_{m}(t)\right]}\,
 L_{mi}(t)\, dt\right)=\Delta^{(n),MAX} \,\cdot\, \mathcal{A}
\end{equation}
Therefore, convergence is guaranteed if $\mathcal{A}<1$ as 
 $\Delta^{(n+1),MAX}<\Delta^{(1),MAX}\,\mathcal{A}^{n}$. Obviously convergence occurs
also for less restrictive conditions but this is harder to constrain.
In our computations we find $\mathcal{A}<1$
for almost all wavelengths and depth positions.
Only at some specific locations in the ($A_{V}^\lambda$) grid (those associated
with some line wings), $\mathcal{A}$ can take values $<10$.
However, a close look at successive variations of $\Delta_{m}^{(n+1)}-\Delta_{m}^{(n)}$
at those locations shows that $\Delta_{m}^{(n+1)}-\Delta_{m}^{(n)}$
 is effectively null after a few iterations.

\begin{figure}[hb]
    \centering \includegraphics[width=6.2cm,angle=-90]{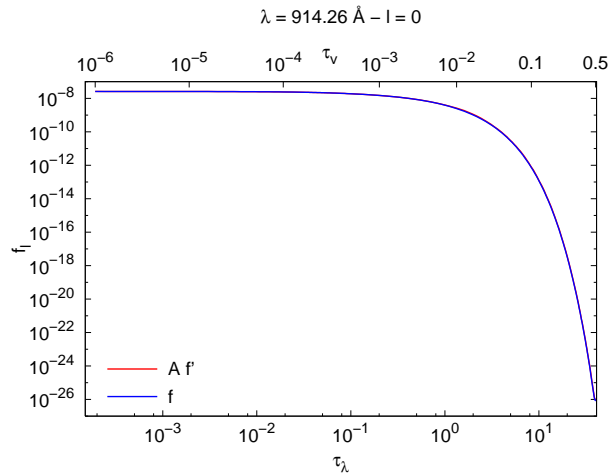}
    \caption{Comparison of $\mathbf{f}$ and $\mathbf{A}\,\mathbf{f'}$ for
     $l$=0 and $\lambda=914.26\,\textrm{\AA}$. The abscissa corresponds to
     $\tau_{v}$ for the upper scale and to $\tau_{line}$ for the lower scale.}
    \label{fig:appendix-demo}
\end{figure}

A final test to validate our numerical solution is to compute
the numerical derivative of our solution and compare  $\mathbf{f}$ 
with $\mathbf{A}\,\mathbf{f'}$ (see Eq.~(\ref{eq-system2})).
Although grain properties are kept uniform, inclusion of gas absorption
makes  $\mathbf{R}'(\tau)\neq0$ and thus
$\mathbf{L}'(\tau)\neq0$ in Eq.~(\ref{eq-new-eq}). 
Figure~\ref{fig:appendix-demo} shows a typical example for a test cloud with $A_V=1$ and
$n_{H}=300\,{\rm cm}^{-3}$, illuminated by the standard radiation field
on both sides. In particular, we compare the  $f_l$ ($l$=0) component of
$\mathbf{f}$ with $\mathbf{A}\,\mathbf{f}'$ at 
$\lambda=914.26\,\textrm{\AA}$,  a ${\rm H}_{2}$ line wing  with a total optical depth of 80. 
Hence, variations of physical conditions along the cloud are large.
It can be seen that the agreement is excellent. 
In a continuum "free of lines" wavelength range, agreement is perfect, and there is nothing to show.
Hence, the derived numerical solution 
is a very good approximated solution to the radiative transfer problem. 

\section{Eigenvalues and eigenvectors of $\mathbf{A^{-1}}(\tau)$\label{ap_diag}}

We describe here our method to compute the eigenvalues and 
eigenvectors of the $\mathbf{A^{-1}}(\tau)$ matrix (see Eq.(\ref{eq-Asis})).
Note that $\mathbf{A}(\tau)$ and $\mathbf{A}^{-1}(\tau)$ have the same
eigenvectors, but  $k_{m}^{-1}(\tau)$ and $k_{m}(\tau)$ eigenvalues respectively.

A first trick is to turn this diagonalization problem to a symmetric problem.
Let us call $\mathbf{R}_{m}(\tau)$  an eigenvector
of $\mathbf{A}(\tau)$ with  $R_{lm}(\tau)$ components and $k_{m}^{-1}(\tau)$ eigenvalues.
Thus, $\mathbf{R}(\tau)$ is the matrix formed by the  $\mathbf{R}_{m}(\tau)$ eigenvectors
and we can write:\[
\left(\begin{array}{cccccc}
0 & h_{0}^{-1} & . & . & . & .\\
h_{1}^{-1} & 0 & 2h_{1}^{-1} & . & . & .\\
. & 2h_{2}^{-1} & 0 & \ddots & . & .\\
. & . & 3h_{3}^{-1} & \ddots & (L-1)h_{L-2}^{-1} & .\\
. & . & . & \ddots & 0 & Lh_{L-1}^{-1}\\
. & . & . & . & Lh_{L}^{-1} & 0\end{array}\right).\left(\begin{array}{c}
R_{0m}\\
R_{1m}\\
R_{2m}\\
\vdots\\
R_{L-1,m}\\
R_{Lm}\end{array}\right)=k_{m}^{-1}(\tau)\,\left(\begin{array}{c}
R_{0m}\\
R_{1m}\\
R_{2m}\\
\vdots\\
R_{L-1,m}\\
R_{Lm}\end{array}\right)\]
with $h_{l}(\tau)=(2l+1)\,(1-\omega(\tau)\sigma_{l}(\tau))$. If we now define 
$\mathbf{G}(\tau)$ as diagonal matrix with $g_{ll}(\tau)=h_{l}^{1/2}(\tau)$, left-multiplication
of  the previous equation by $\mathbf{G}(\tau)$ and insertion of the identity matrix
${\rm I}=\mathbf{G}^{-1}(\tau)\,\mathbf{G}(\tau)$ between
$\mathbf{A}(\tau)$ and $\mathbf{R}(\tau)$ gives\footnote{Left-multiplication
 by a diagonal matrix multiplies rows by a constant, and right multiplication 
multiplies columns.}:
\begin{equation}
\left(\begin{array}{cccccc}
0 & \frac{1}{\sqrt{h_{0}h_{1}}} & . & . & . & .\\
\frac{1}{\sqrt{h_{0}h_{1}}} & 0 & \frac{2}{\sqrt{h_{1}h_{2}}} & . & . & .\\
. & \frac{2}{\sqrt{h_{1}h_{2}}} & 0 & \ddots & . & .\\
. & . & \frac{3}{\sqrt{h_{2}h_{3}}} & \ddots & \frac{L-1}{\sqrt{h_{L-2}h_{L-1}}} & .\\
. & . & . & \ddots & 0 & \frac{L}{\sqrt{h_{L-1}h_{L}}}\\
. & . & . & . & \frac{L}{\sqrt{h_{L-1}h_{L}}} & 0\end{array}\right).\left(\begin{array}{c}
h_{0}^{1/2}R_{0m}\\
h_{1}^{1/2}R_{1m}\\
h_{2}^{1/2}R_{2m}\\
\vdots\\
h_{L-1}^{1/2}R_{L-1,m}\\
h_{L}^{1/2}R_{Lm}\end{array}\right)=k_{m}^{-1}(\tau)\,\left(\begin{array}{c}
h_{0}^{1/2}R_{0m}\\
h_{1}^{1/2}R_{1m}\\
h_{2}^{1/2}R_{2m}\\
\vdots\\
h_{L-1}^{1/2}R_{L-1,m}\\
h_{L}^{1/2}R_{Lm}\end{array}\right)
\end{equation}
This new symmetric matrix is called $\mathbf{\tilde{A}}(\tau)$, and 
$\mathbf{\tilde{R}}(\tau)$ is the matrix of its eigenvectors. 
The $\mathbf{\tilde{A}}(\tau)$ matrix has the same eigenvalues
$k_{m}^{-1}(\tau)$  as $\mathbf{A}(\tau)$, although the eigenvectors $\mathbf{R}(\tau)$
and $\mathbf{\tilde{R}}(\tau)$ are different but related by
 $\mathbf{\tilde{R}}(\tau)=\mathbf{G}(\tau)\,\mathbf{R}(\tau)$.
These symmetric matrixes are easier to diagonalize numerically. 
Eigenvectors are computed by the recurrence relation:
\[
R_{0m}(\tau)=1\]
\[
R_{1m}(\tau)=(1-\omega(\tau))/k_{m}(\tau)\]
\begin{equation}
R_{lm}(\tau)=\frac{1}{lk_{m}(\tau)}\,\left[h_{l-1}(\tau)\, R_{l-1,m}(\tau)-(l-1)\, 
k_{m}(\tau)\, R_{l-2,m}(\tau)\right]
\label{eq-Rrecu}
\end{equation}
where, compared to Roberge (1983), $\omega(\tau)$ is a $\tau$--dependent effective
albedo including line absorption.

\section{Inverse and derivative of $\mathbf{R}(\tau)$\label{ap_invr}}

Here we show how $\mathbf{R}^{-1}(\tau)$ is computed. Unfortunately, $\mathbf{A}(\tau)$
is not a symmetric matrix, so that $\mathbf{R}^{-1}(\tau)\neq\mathbf{R}^{T}(\tau)$. 
However, we can apply the same method as above to turn 
$\mathbf{R}^{-1}(\tau)$ into $\mathbf{\tilde{R}}^{T}(\tau)$.
Since $\mathbf{\tilde{A}}(\tau)$ is symmetric,
the matrix formed with its eigenvectors is orthogonal. Thus, 
using the same notations, we have:

\begin{equation}
\mathbf{\tilde{R}}^{T}(\tau)\,\mathbf{\tilde{R}}(\tau)=\mathbf{J}(\tau)=(\mathbf{G}(\tau)
\,\mathbf{R}(\tau))^{T}\,\mathbf{G}(\tau)\,\mathbf{R}(\tau)
\end{equation}
where $\mathbf{J}(\tau)$ is a diagonal matrix with 
$J_{ll}(\tau)=\left\Vert \mathbf{\tilde{R}}_{l}(\tau)\right\Vert ^{2}$ elements.
Hence:
\begin{equation}
\mathbf{R}^{-1}(\tau)=\mathbf{J}^{-1}(\tau)\,\mathbf{R}^{T}(\tau)\,\mathbf{G}^{2}(\tau)
\end{equation}

The inclusion of the depth dependence in the spherical harmonics method unfortunately
forces to calculate the $\mathbf{R}(\tau)$ derivative 
respect to $\tau$.  Ideally, we could start to derivate  the
$R_{lm}(\tau)$ recurrence relations shown in Eq~(\ref{eq-Rrecu}) to get:
\[
R'_{0m}=0\]
\[
R'_{1m}=-\frac{\omega'k_{m}+(1-\omega)k'_{m}}{k_{m}^{2}}\]
\begin{equation}
R'_{lm}=-\frac{k'_{m}}{lk_{m}^{2}}\,\left[h_{l-1}R_{l-1,m}-(l-1)k_{m}R_{l-2,m}\right]
+\frac{1}{lk_{m}}\,\left[h'_{l-1}R_{l-1,m}+h_{l-1}R'_{l-1,m}-(l-1)
\left(k'_{m}R_{l-2,m}+k_{m}R'_{l-2,m}\right)\right]
\end{equation}
with 
\begin{equation}
h'_{l}(\tau)=-(2l+1)\,[\omega'(\tau)\sigma_{l}(\tau)\,+\omega(\tau)\sigma_{l}'(\tau)]
\end{equation}
Unfortunately, $\omega'$, $\sigma_{l}'$ and $k'_{m}$ have to be computed also numerically,
which is quite unstable in the most external cloud positions
due to the large variations of $\tau_{\lambda}$
at line wing wavelengths (where the line opacity becomes comparable to
the dust opacity) compared to deeper inside the cloud where $\tau_{\lambda}$
at the same wavelength becomes saturated
(the dust opacity becomes insignificant respect to the line opacity).
Besides, a symmetric difference scheme does not provide satisfactory results
because   $\frac{\omega_{n+1}-\omega_{n-1}}{\tau_{n+1}-\tau_{n-1}}$ only
gives an approximation to $\omega'$ at $\tau=\frac{\tau_{n-1}+\tau_{n+1}}{2}$
which, in general, is not $\tau_{n}$.
We solved this problem by derivating directly the computed values of
$R(\tau)$. To avoid  irregular steps in $\tau$,  a second order polynomial was
fit to $R_{lm}(\tau_{i-2})$, $R_{lm}(\tau_{i})$ and $R_{lm}(\tau_{i+2})$, and 
the value of its analytical derivative was then used. 
The resulting derivative $\mathbf{R}'(\tau)$ is smooth enough
to be applied in the numerical computation.

\section{Mean radiation field intensity\label{ap_intensity}}

In section~\ref{sub_Mean-intensity} we deduced the simple form that
the mean intensity takes in  the spherical harmonics method,
 i.e. $J(\tau)=f_{0}(\tau)$. 
However, in some cases of astrophysical
interest (e.g. a two sides asymmetrically illuminated cloud) one needs to distinguish
 the fraction of  radiation field  coming from each side of the cloud. 
In this case, two half sums have to be computed. Here we give the analytical 
expressions that we use to compute $J^{\pm}(\tau)$.
For radiation coming from the $\tau=0$ side we have:
\begin{equation}
J^{-}(\tau)=\frac{1}{2}\int_{-1}^{0}I(\tau,\mu)\, 
d\mu=\frac{1}{2}\,\sum_{l}(2l+1)\, f_{l}(\tau)\,\int_{-1}^{0}P_{l}(\mu)\, d\mu
\label{eq:_J-}
\end{equation}
And for radiation coming from the $\tau=\tau_{max}$ side we have:
\begin{equation}
J^{+}(\tau)=\frac{1}{2}\int_{0}^{+1}I(\tau,\mu)\, d\mu=\frac{1}{2}\,
\sum_{l}(2l+1)\, f_{l}(\tau)\,\int_{0}^{+1}P_{l}(\mu)\, d\mu
\label{eq:_J+}
\end{equation}
If we define $\mathcal{Q}_{l}=\int_{0}^{+1}P_{l}(\mu)\, d\mu$,
with:
\begin{equation}
\mathcal{Q}_{l}=\left\{ \begin{array}{lc}
1 & l=0\\
0 & l\,\,\textrm{even and }>0\\
\frac{P_{l-1}(0)}{l+1} & l\,\,\textrm{odd}\end{array}\right.
\end{equation}
parity gives $\int_{-1}^{0}P_{l}(\mu)\, d\mu=(-1)^{l}\,\mathcal{Q}_{l}=
-\mathcal{Q}_{l}$ (using $\mathcal{Q}_{l}=0$ for $l$ even).
Inserting Eq.~(\ref{eq-final_f}) in Eqs~(\ref{eq:_J-}) and (\ref{eq:_J+}) we get:
\begin{equation}
J^{-}(\tau)=\frac{1}{2}\sum_{m=-M}^{-1}\left(C_{m}\, 
E_{m}^{-}(\tau)-D_{m}^{-}(\tau)\right)\,\left(1-\sum_{l=1}^{L}(2l+1)\,
\mathcal{Q}_{l}\, R_{lm}(\tau)\right)
+\frac{1}{2}\sum_{m=1}^{M}\left(C_{m}\, 
E_{m}^{+}(\tau)+D_{m}^{+}(\tau)\right)\,\left(1-\sum_{l=1}^{L}(2l+1)\,
\mathcal{Q}_{l}\, R_{lm}(\tau)\right)
\end{equation}
\begin{equation}
J^{+}(\tau)=\frac{1}{2}\sum_{m=-M}^{-1}\left(C_{m}\, 
E_{m}^{-}(\tau)-D_{m}^{-}(\tau)\right)\,\left(1+\sum_{l=1}^{L}(2l+1)\,
\mathcal{Q}_{l}\, R_{l,m}(\tau)\right)
+\frac{1}{2}\sum_{m=1}^{M}\left(C_{m}\, 
E_{m}^{+}(\tau)+D_{m}^{+}(\tau)\right)\,\left(1+\sum_{l=1}^{L}(2l+1)\,
\mathcal{Q}_{l}\, R_{lm}(\tau)\right)
\end{equation}
Taking into account the fact that $\mathcal{Q}_{l}=0$ if $l$ is
even, and $R_{l,-m}=-R_{lm}$ if $l$ is odd, we now  define (for $m>0$)
\begin{equation}
\mathcal{S}_{m}(\tau)=\sum_{l\, odd}(2l+1)\,\mathcal{Q}_{l}\, R_{l,m}(\tau)
\end{equation}
to write:
\begin{equation}
J^{-}(\tau)=\frac{1}{2}\sum_{m=-M}^{-1}\left(C_{m}\, E_{m}^{-}(\tau)-D_{m}^{-}
(\tau)\right)\,(1+\mathcal{S}_{m}(\tau))
+\frac{1}{2}\sum_{m=1}^{M}\left(C_{m}\, E_{m}^{+}(\tau)+D_{m}^{+}(\tau)\right)
\,(1-\mathcal{S}_{m}(\tau))
\end{equation}
\begin{equation}
J^{+}(\tau)=\frac{1}{2}\sum_{m=-M}^{-1}\left(C_{m}\, 
E_{m}^{-}(\tau)-D_{m}^{-}(\tau)\right)\,(1-\mathcal{S}_{m}(\tau))
+\frac{1}{2}\sum_{m=1}^{M}\left(C_{m}\, E_{m}^{+}(\tau)+D_{m}^{+}(\tau)\right)
\,(1+\mathcal{S}_{m}(\tau))
\end{equation}
Therefore, the fraction of the mean intensity coming from each side of the cloud
can be easily determined at each depth.
The resulting $J^{\pm}(\tau)$ values can then be used to evaluate the  escape
probably  of any FUV photon emitted within the cloud.

\end{appendix}

\end{document}